\documentclass{aa}  
\usepackage{amsmath}
\usepackage{graphicx}
\usepackage{txfonts}
\usepackage{siunitx}
\newcommand{\BGF}[1]{{\bf #1}}
\newcommand{\BGS}[1]{{\normalfont #1}}
\newcommand{\BG}[1]{{\normalfont #1}}
\usepackage{natbib,twoopt}
\bibpunct{(}{)}{;}{a}{}{,}             
\makeatletter
  \newcommandtwoopt{\citeads}[3][][]{\href{http://adsabs.harvard.edu/abs/#3}%
    {\def\hyper@linkstart##1##2{}%
     \let\hyper@linkend\@empty\citealp[#1][#2]{#3}}}
  \newcommandtwoopt{\citepads}[3][][]{\href{http://adsabs.harvard.edu/abs/#3}%
    {\def\hyper@linkstart##1##2{}%
     \let\hyper@linkend\@empty\citep[#1][#2]{#3}}}
  \newcommandtwoopt{\citetads}[3][][]{\href{http://adsabs.harvard.edu/abs/#3}%
    {\def\hyper@linkstart##1##2{}%
     \let\hyper@linkend\@empty\citet[#1][#2]{#3}}}
  \newcommandtwoopt{\citeyearads}[3][][]%
    {\href{http://adsabs.harvard.edu/abs/#3}
    {\def\hyper@linkstart##1##2{}%
     \let\hyper@linkend\@empty\citeyear[#1][#2]{#3}}}
\makeatother
\begin{document} 

   \title{A new compact young moving group around V1062 Sco} 
   
   \author{Siegfried R\"oser
          \inst{1,2}
          \and
          Elena Schilbach
          \inst{1,2}
          \and
          Bertrand Goldman
          \inst{3,4}
          \and
          Thomas Henning
          \inst{3}
          \and
          Attila Moor
          \inst{5}
          \and
          Aliz Derekas
          \inst{5,6}
          }

\institute
{Zentrum f\"ur Astronomie der Universit\"at
Heidelberg, Landessternwarte, K\"{o}nigstuhl 12, 69117 Heidelberg, Germany
\and
Zentrum f\"ur Astronomie der Universit\"at
Heidelberg, M\"{o}nchhofstra\ss{}e 12-14, 69120 Heidelberg, Germany\\
    \email{roeser@ari.uni-heidelberg.de, elena@ari.uni-heidelberg.de}
\and
Max-Planck-Institut f\"ur Astronomie, K\"onigstuhl 17, 69117 Heidelberg, Germany
\email{goldman@mpia.de}
\and 
Observatoire astronomique de Strasbourg, Universit\'e de Strasbourg - CNRS UMR 7550,
 11 rue de l'Universit\'e, 67000, Strasbourg, France
\and
Konkoly Observatory, Research Centre for Astronomy and Earth Sciences, Hungarian Academy of Sciences, H-1121 Budapest, Konkoly Thege Mikl\'os \'ut 15-17, Hungary
\and
ELTE E\"otv\"os Lor\'and University, Gothard Astrophysical Observatory, Szombathely, Hungary
}

\date{Received 30 October 2017; accepted 27 December 2017}


  \abstract
    {}
   {We are searching for new open clusters or moving groups in the Solar neighbourhood.} 
{We used the Gaia-TGAS catalogue, cut it into narrow proper motion and parallax slices and searched for significant spatial over-densities of stars in each slice. We then examined stars \BG{forming} 
over-densities in optical and near-infrared colour-magnitude
diagrams to determine if they are compatible with isochrones of a cluster.}
   {We detected a hitherto unknown moving group or cluster in the Upper Centaurus Lupus (UCL) section of the Sco-Cen OB-association at a distance of 175 pc from the Sun. It is a group of \BG{63} co-moving stars with ages of less than 10 to about 25~Myr. For the brightest stars, 
which are present in the Gaia-TGAS catalogue the mean difference between kinematic and trigonometric distance moduli is $-0.01$~mag with a standard deviation of 0.11~mag.
Fainter cluster candidates are found in the HSOY catalog, where no trigonometric parallaxes are available. For a subset of \BG{our} 
candidate stars, we obtained radial velocity measurements at the MPG/ESO 2.2-metre telescope in La Silla. Altogether we found twelve members with confirmed radial velocities and parallaxes, 31 with parallaxes or radial velocities, and 20 candidates from the
convergent point method. The \BG{isochrone} masses of our 63 members range from 2.6 $\rm{M}_\odot$ to 0.7 $\rm{M}_\odot$.}
   {}


   \maketitle

\section{Introduction}

For more than a century, the Scorpius-Centaurus OB-association (Sco-Cen) has been 
studied as a moving group of early-type stars. \citet{1964ARA&A...2..213B}
divided the huge nearby association into three subgroups, Upper Scorpius,
Upper Centaurus Lupus (UCL), also known as Sco~OB2\verb+_+3, and
Lower Centaurus Crux. After ESA's Hipparcos mission, 
accurate proper motions and trigonometric parallaxes became
available for many early-type stars in Sco-Cen.
This material has been used by \citet{1999AJ....117..354D} for their
\BG{fundamental} paper on OB-associations.

Among the Hipparcos objects in the UCL subgroup of Sco-Cen, \citet{1999AJ....117..354D} identified
221 members in this field, with a
mean distance of 140 $\pm$ 2 pc, 66 B, 68 A, 55 F, 25 G, six K,
and one M-type star. 
They also noted that the star-forming regions associated with the Lupus
clouds contained several tens of low-mass pre-main-sequence stars with estimated ages of
$\approx 3$ Myr \citep[see the references in][]{1999AJ....117..354D}. 
These star-forming regions (SFR) cover most of UCL, except a small area
in the south-eastern corner of UCL. It is in this area where we found a compact,
young moving group of stars \BGS{while searching for open clusters in the TGAS catalogue \citep{2015A&A...574A.115M,2016A&A...595A...4L}.

\BGS{The first Gaia data release \citep[Gaia-DR1, ][]{2016A&A...595A...2G}, and especially the publication of TGAS  with its high-precision proper motions and trigonometric parallaxes fostered work on open clusters and moving groups by a number of authors. Without aiming for completeness we mention here the work performed by \citet{2017A&A...601A..19G} on nearby open clusters and of \citet{2017arXiv171107699R} who, combining information from Gaia-DR1 and the Gaia-ESO Survey \citep{2012Msngr.147...25G} data, derive new distances and ages of eight nearby open clusters. \citet{2017AJ....153..257O} used the TGAS catalogue to search for high-confidence comoving pairs of stars. In doing so they find some ten thousand comoving star pairs with separations as large as 10 pc. They also find larger groups corresponding to known open clusters and OB associations and several new comoving groups.}

Studying the structure and the behaviour of moving groups may shed light on the process of star formation in associations. We are  
guided by the statement by Richard Larson \citep{2002ASPC..285..442L} in his summary during the conference "Modes of Star Formation and the Origin of Field Populations":
\em{"When associations disperse, they retain some kinematic coherence because their
internal motions are relatively small, and they then become moving groups or
streams. The weak-lined T Tauri stars found by X-ray observations that were
discussed at this meeting probably represent some of the remnants of associations
that are just now dissolving into the field and becoming moving groups."}}
\BGS{So, it was worthwhile to study our new moving group in more detail as all ingredients discussed by Larson are in place, an association, a good number of T Tauri and weak-line T Tauri stars, and a moving group.}

In this paper we describe in Section 2  how we detected this group, as well as the spectroscopic follow-up measurements we carried out with the FEROS instrument at the MPG/ESO~2.2~m~Telescope in La~Silla
to obtain radial velocities and stellar parameters.
In Section 3 \BG{we provide} a discussion of the properties
of the newly found group. Then, in Section 4 we compare our new group with 
the star-forming areas in the other parts of UCL, and a short summary concludes the paper.

\section{Detection of the V1062 Sco moving group}
\subsection{Astrometric detection}\label{astdet}
After the first Gaia data release in September 2016, we performed a cluster search programme
similar to the one described in \citet{2016A&A...595A..22R}, but with TGAS replacing
the Tycurat catalogue which we used originally. In brief the method consists in cutting the TGAS catalogue into slices as follows: in proper motion space we place a grid with grid points $(i,k)$ at each integer value (in mas\BGS{/y}) from $-50$\,mas/y to +50\,mas/y in both proper motion components. A slice at $(i,k)$ contains all stars for which $(\mu_{\alpha}\cos{\delta} -i)^2+(\mu_{\delta}-k)^2 \leq 1.5^2 \, \rm (mas/y)^2$ holds. For each  slice,  we then check the distribution of its stars over the sky and 
search for possible overdensities which can be considered as candidates of clusters or moving groups. 
  
  \begin{figure}[h!]
   \centering
   \includegraphics[width=0.48\textwidth]{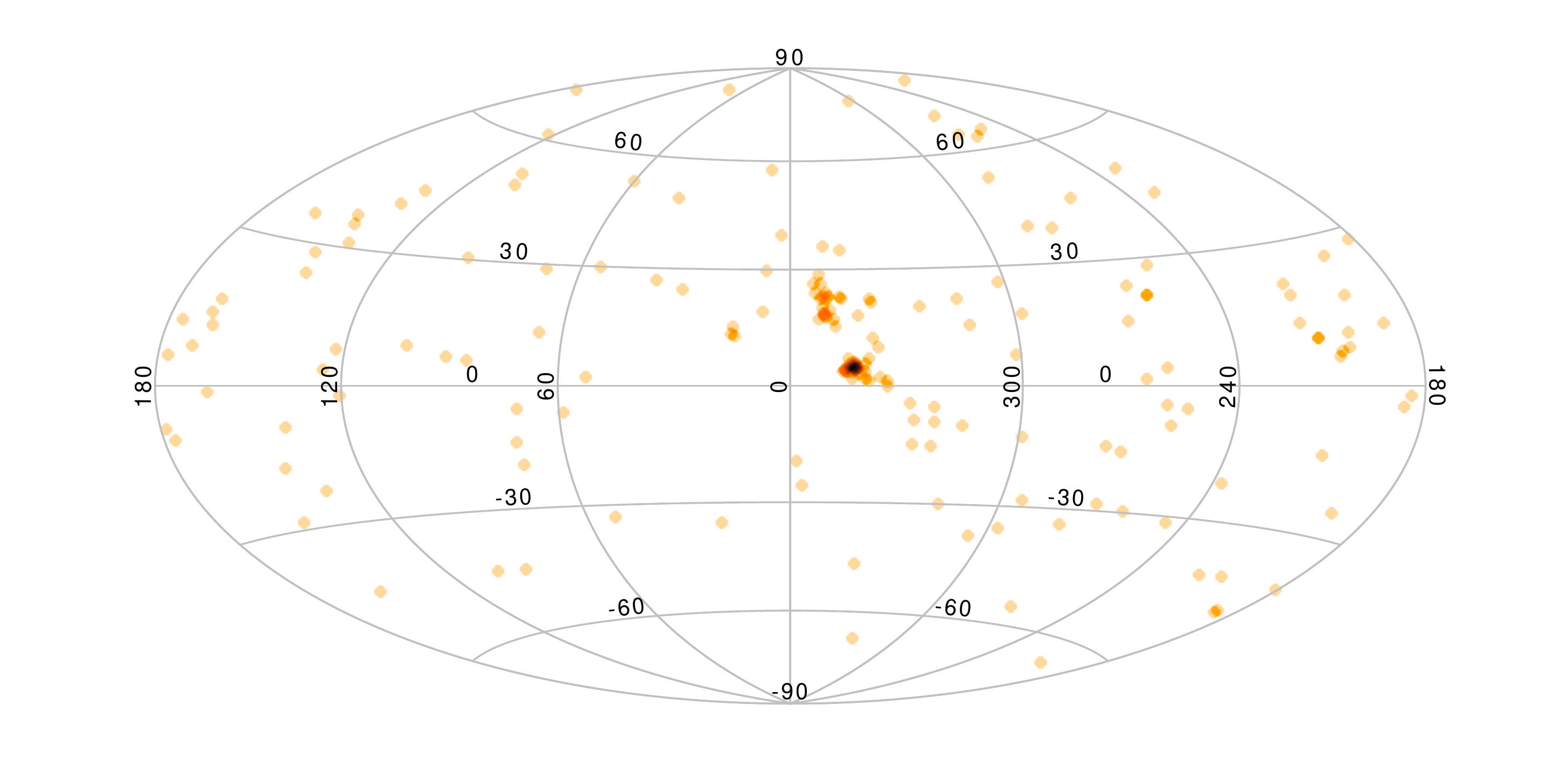}
      \caption{Sky map (in Galactic coordinates) showing the stars from the TGAS catalog with parallaxes between 5 and 10~mas and in the
      proper motion slice around $(\mu_{\alpha}\cos{\delta},\mu_{\delta}) = (-12,-22)$ mas/y. For further information on the selection see text.}
         \label{TWA}
   \end{figure}

\begin{figure}[h!]
   \centering
   \includegraphics[width=0.48\textwidth]{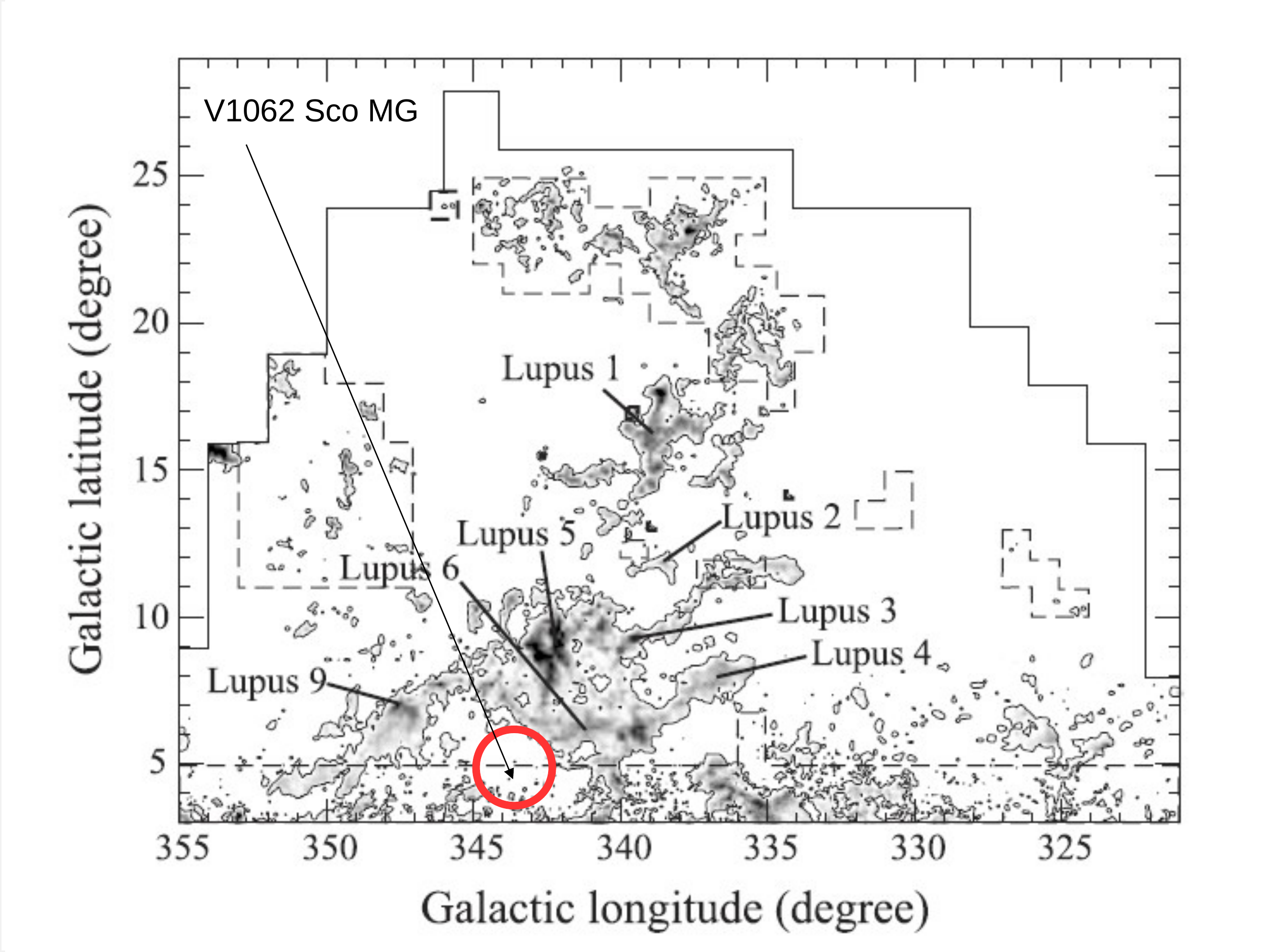}
      \caption{The location of the V1062 Sco Moving Group \BGS{(red circle)} overlayed to the $^{12}$CO map from \citet{2001PASJ...53.1081T}. \BGS{The radius of the circle is 2.75\degr  or 8.4 pc at a distance of 175 pc and roughly  corresponds to the extent of the group.}}
         \label{comap}
   \end{figure}

The \BG{availability} of the TGAS parallaxes allows us to improve the effectivity of the method by cutting the catalogue into parallax slices, too.
Since we are especially interested in finding unknown clusters in the vicinity of the Sun, we consider a parallax slice 
between 5 mas and 10 mas which contains the stars between 100 and 200 pc from the Sun. Compared to the full TGAS sky with some 2 million stars this parallax slice only includes 160,000 stars. The latter is then cut into proper motion slices. In Fig.~\ref{TWA} we display a sky map in Galactic coordinates of the proper motion slice $(i,k) = (-12,-22)$ mas/y which only contains 159 stars.
Here we found a hitherto unknown overdensity containing 25 stars (of 159) in a radius of 2.5 degrees around $(l,b) = (343\fdg6,+4\fdg3)$ in UCL. The fainter filamentary structure north of the overdensity may also be a substructure in Sco-Cen, but has not yet been investigated by us.
Our newly found overdensity in the south-eastern corner of UCL occupies an area which is void on the $^{12}$CO map from \citet{2001PASJ...53.1081T}. We show its location in detail on Fig.~\ref{comap}. Also, the trigonometric parallaxes of these 25 stars show a narrow range between 5.3 and 6.3~mas in TGAS (cf. Fig\ref{parallaxes}), we henceforth refer to this overdensity of co-moving stars as the V1062 Sco Moving Group (V1062 Sco MG) according to its prominent member V1062 Sco, postponing the discussion if V1062 Sco is a true member or not to Section \ref{ind_stars}. \BGS{Parallel to our finding of this new moving group this object has been detected in the above-mentioned paper by  \citet{2017AJ....153..257O} as their "Group 11", but without further discussion.  }
\subsection{Radial velocities and stellar parameters}
In order to get additional confirmation for the existence of the V1062 Sco MG, we carried out FEROS observations at the MPG/ESO~2.2~m~Telescope in July 2017
and determined radial velocities ($RV$) for a subset of possible moving group candidates.
FEROS is a fiber-fed, high-resolution (48,000) cross-dispersed optical spectrograph mounted on the MPG/ESO 2.2-metre telescope in La~Silla \citep{Kaufe99}.
FEROS has two set-ups: one of its two fibers points to the science target, and the second points either to a calibration lamp for better wavelength calibration, or to the sky.
For brighter targets, we used both set-ups, for fainter targets only the sky set-up.
We found that for our observations, the wavelength calibration errors are a minor contributor to the error budget. 
The observing log is given in Table~\ref{FEROSlog}.

To reduce the spectra we used the CERES pipeline \citep{Brahm17}, a reduction and extraction pipeline dedicated to {\'e}chelle spectrographs.
In a nutshell, for each of our three observing nights, CERES identifies the calibration files and obtains a master bias and dark frames through median filtering. The flat field is not removed at this stage but instead the blaze function obtained from continuum lamp frames is divided after extraction, which also corrects the column-averaged pixel-to-pixel variations of the CCD. It then identifies the order traces and fits them with a high-order polynomial. It measures the scattered light in the inter-order zones, interpolating over the spectra and performs a 2-D median filtering of the resulting map.
After subtraction of the scattered light, the spectra are optimally extracted. 
The spectra are wavelength-calibrated using the ThAr lamp spectra, observed either simultaneously with the target, or during the afternoon.

CERES measures radial velocities by cross-correlating the spectra with binary masks indicating the narrow absorption lines, following \citet{Mayor03}. As CERES is optimized for late-type stars in its standard configuration, it provides masks corresponding to G2, K5 and M2 spectral types.
Measurements of early-type stars is notoriously difficult as their spectra exhibit few absorption lines which are considerably broadened by stellar rotation. 
In the following we deal only with spectra of candidates with effective temperatures smaller than 7000\,K.
The uncertainties on the $RV$ is determined as a function of the signal-to-noise ratio (SNR) at the Mg-triplet wavelength and the width of the \BG{cross-correlation function (CCF)}, with parameters determined from a Monte-Carlo simulation using degraded high SNR spectra.
The measurements are reported in Table~\ref{FEROSlog}.
\BGS{For targets with multiple observations we report the weighted average and the error on the mean of all the available measurements and  list these stars in Table~\ref{FEROSmean}. The dispersion of these individual observations roughly confirms the uncertainties derived from the SNRs.}

To obtain the stellar parameters, we applied the ZASPE \BGS{\citep{Brahm16}} pipeline. 
The pipeline is optimized to use the spectral library of \citet{Brahm16}, created in order to reduce the biases in the stellar parameters with respect to the measurements of SWEET--Cat \citep{Santo13}.
This library covers a range of stellar temperatures up to 7000\,K.  
\BGS{The precision of the parameters} can be estimated from the repeated observations of some targets, typically a few 0.1\,dex in $\log\,g$ , 0.1\,dex in [Fe/H], 200\,m/s in $v\sin i$, a few dozens Kelvin in $T_{\rm eff}$. Average values of these parameters are given in Table~\ref{FEROSmean}.
Our cluster members or candidates (m(4d) to m(6d), see Section~\ref{mship}) have a mean metallicity of 0.04\,dex, with a spread of 0.1\,dex. Their rotational velocity ranges from 5 to 80\,km/s. 
\BGS{Only one cluster candidate member shows H$_\alpha$ emission, namely} star 25, which also displays (stronger) absorption. This general lack of emission, indicating no accretion, is consistent with the age range determined from the isochrones (see Section~\ref{CMD}).
%
\subsection{Phase space coordinates}
\BGS{For eleven out of the 25 kinematic candidate stars mentioned in Sec. \ref{astdet} we obtained  well-determined radial velocities with formal errors less than 1~km/s. Their $RV$ pile up between  1.0 km/s to +3.6 km/s and give $RV_{mean} = +2.24 \pm 0.85~$ km/s.} 
Moreover, their parallaxes turned out to be within a narrow range between 5.4 mas and 6.2 mas, corresponding
to a mean distance from the Sun of $175\pm7$~pc. For these stars we now have the complete set of 6-D phase 
space coordinates based on direct observations. \BGS{Hence we compute the mean space coordinates and velocities of the group in barycentric Galactic coordinates to be}:
\begin{small}
\begin{equation}
(X,Y,Z,U,V,W) = (167.20,-49.14,13.44,-3.80,-19.96,-4.06).
\label{finalxyz}
\end{equation}
\end{small}
\BGS{Here the X-axis points to the Galactic centre, the Y-axis to the direction of Galactic rotation and the Z-axis points to the Galactic North Pole.  $U, V, W$ are the corresponding components of the space velocity.}
The standard deviation in the 3 components of the space velocity is rather small
$(\sigma_U,\sigma_V,\sigma_W) = (0.7,0.9,0.3)$ km/s. \BGS{The individual residuals in velocity space do not show systematic effects such as rotation or expansion.
The standard deviation sets an upper limit of 1 km/s for the one-dimensional velocity dispersion of the group, and we will show in the discussion in Sec. \ref{mship} that this is large compared to the formal error at least for that component of the tangential velocity which is perpendicular to the direction to the convergent point.}
\section{Properties the V1062 Sco Moving Group}
We can use the phase space coordinates (\ref{finalxyz}) of the V1062 Sco moving group to perform a search of its members. This can be done on the basis of the convergent point (CP) method we have applied \BG{to find} 
members in the Hyades cluster \citep{hyades11,2013A&A...559A..43G}. In case of the V1062 Sco moving group, we extended the search to a distance of 10 pc around the centre $(X,Y,Z) = (167.20,-49.14,13.44)$ from (\ref{finalxyz}).
Using the proper motions given in TGAS, the method predicts kinematic parallaxes $\varpi_{kin}$, as well as the velocity components in the tangential plane parallel and perpendicular to the direction of the convergent point. In the ideal case of an exactly parallel space  motion the perpendicular component has to be zero. Taking into account a possible internal velocity dispersion and the accuracy of the input data, we allowed a maximum of  3~km/s for the perpendicular velocity component. With these restrictions we found 35 TGAS stars as kinematic candidates of the moving group. 
Further we required that the difference between the predicted 
kinematic parallaxes $\varpi_{kin}$ and the observed trigonometric parallaxes $\varpi_{trig}$ from TGAS \BG{to be} smaller than the mean error of this difference, i.e.
\begin{equation} 
| \varpi_{kin} - \varpi_{trig} | < (\sigma_{\varpi_{kin}}^2+\sigma_{\varpi_{trig}}^2+0.3^2)^{1/2}, 
\label{dplx}
\end{equation}
where 0.3 mas was taken as a possible systematic error of trigonometric parallaxes \citep{2016A&A...595A...4L}.
The application of the restriction~(\ref{dplx}) deletes two candidates, so
we finally retained 33 kinematic members in our group confirmed by their trigonometric parallaxes.
We  publish them in Table~\ref{T} available as on-line material together with the paper.

In order to find fainter candidates of the V1062 Sco moving group, we also used the HSOY catalogue \citep{2017A&A...600L...4A}.
HSOY was constructed by \BGS{using Gaia-DR1 positions as an additional epoch together with \mbox{PPMXL} positions and proper motions to obtain formally improved
proper motions}. Although HSOY may have significant systematic proper motion errors over the sky, these should be unimportant for work in small areas of some tens of square degrees. We restricted our search within HSOY to \BGS{G $\leq$ 14 mag} that yielded  17 million stars; this subset is called HSOY14 in the following.
This restriction was driven by the fact that the moving group is located south of
declination -30 degrees and the quality of the proper motions in HSOY for fainter stars deteriorates in this quarter of the sky.
Using the convergent point method with the phase space coordinates~(\ref{finalxyz}) we derived kinematic \BGS{parallaxes}, which, however, cannot yet be verified by the trigonometric parallax observations.
\BGS{So, as in the case of the TGAS stars, the distance moduli and absolute magnitudes are derived from kinematic parallaxes.
The CP method yielded 149 candidates. We then usually discard background stars which appear below the ZAMS and only retain stars which fulfil $ M_G \leq 3 \times (G-J) + 2 $ in the $M_G,G-J$ Colour-Magnitude-Diagram (CMD). Fourty stars are discarded by this criterium, all redder than (G-J) =1.0.}
Since HSOY14 contains TGAS as a subset, we had to discard the above mentioned 35 TGAS stars from the \BGS{remaining 109 stars}.
It turns out, however, that from the remaining 74 HSOY14 stars nine were also in the TGAS catalogue. Their proper motions differ slightly in the HSOY14 and TGAS catalogues, and this was a reason why they were not selected as TGAS candidates. Since the formal precision of the proper motions was a factor of almost two better for these stars in HSOY14 compared to TGAS, we kept them as kinematic candidates. Moreover, 7 out of 9 stars fulfilled the condition~\ref{dplx}. These nine stars are given in the
Table~\ref{TH}. 

As the remaining 65 faint HSOY14 candidates were only selected on the basis of their HSOY14 proper motions, we checked the
data quality by cross-matching the stars with UCAC5 \citep{2017AJ....153..166Z} for consistency. \BGS{The reason for doing this is the following: the parent catalogue of HSOY is \mbox{PPMXL} which itself descends from USNO-B1.0 \citep{2003AJ....125..984M}. South of -30\degr declination USNO-B1.0 is of poor quality, as the epoch difference between the first and the second epoch is low in the south, and a possible mismatching leads to grossly erroneous proper motions which may be handed over to HSOY. A comparison with a completely independent source such as UCAC5 removes this problem. So, we only retain those stars whose difference in proper motions between HSOY and UCAC5 is within 1-$\sigma$ of the formal error of this difference}. Thirty-two stars \BGS{from 65} survived this check and are given in Table~\ref{H} in the appendix.
\subsection{Membership}\label{mship}
Due to the selection procedure applied, our sample of kinematic 
candidates includes stars of different membership probability. The quality of membership is based mainly on the observations available for candidates that are consistent with the phase-space coordinates of the V1062 Sco MG defined by (\ref{finalxyz}). 
If, as in the case of HSOY, only four \BGS{parameters}
($\alpha$,\, $\delta$,\, $\mu_{\alpha}$,\, $\mu_{\delta}$) are available and consistent with~(\ref{finalxyz}),
we call these candidates m(4d) members. \BGS{The CP method predicts parallax and radial velocity of candidate members. Both predictions have to be verified by observations. If no trigonometric parallaxes are available we can only make a negative selection by
testing if the predicted absolute magnitudes are compatible with allowed stellar loci in CMDs. By doing this, we indeed  classified a few m(4d) candidates as non-members  due their location in CMDs (see Section~\ref{CMD})}

However, the majority of our candidates had measurements of trigonometric parallaxes ($\varpi_{trig}$) or/and radial velocities ($RV$).  If one of the quantities $RV$, or $\varpi_{trig}$ is measured and consistent with (\ref{finalxyz}), we called the star \BGS{a} m(5d) member since its five coordinates of the 6-D phase space coincide with the assumption of their membership in the moving group V1062 Sco. If $RV$ and  $\varpi_{trig}$ were measured and fulfil (\ref{finalxyz}), we call them m(6d) members. On the other hand, if one of the measurements was inconsistent with (\ref{finalxyz}), we rejected this star as a member and call it non-member (nm). 

In our total sample of 76 kinematic candidates of the V1062 Sco MG we found 12 m(6d) members. These are fully confirmed
members and their existence proves the authenticity of the moving group. Also the 31 m(5d) members are very probable
candidates, as their measured $\varpi_{trig}$  confirms the predicted $\varpi_{kin}$ from the CP method, or, in four cases,
the measured $RV$ confirm the predicted $RV$. The other 20 m(4d) candidates await confirmation by $RV$ and $\varpi_{trig}$
measurements. Finally, in three cases the predicted parallaxes were not confirmed by the observed ones, in one case $RV$ was inconsistent, and one star was rejected due to both inconsistent parallax and radial velocity.
At this point we want to emphasize that, using the photometric criteria, we also identified 8 m(4d) candidates as non-members (see the Section~\ref{CMD}). 
An overview of the membership quality in the  V1062 Sco moving group is given in Table~\ref{mem_stat}. 

The stars in our V1062 Sco MG are remarkably co-moving. Although we have allowed the component of the tangential velocity perpendicular to the direction to the convergent point to $\pm$ 3 km/s, we find an overall standard deviation as small as 1\,km/s around the expectation value of zero. At 175~pc this corresponds to 0.83 mas/y, comparable to the accuracy of the TGAS proper motions. The corresponding scatter of the 12 m(6d) members 
amounts to only 0.3 km/s or less than 0.3 mas/y comparable to the accuracy of the best TGAS stars. As the perpendicular component of the tangential velocity is a measure of the actual 1-d velocity dispersion in a cluster (moving group), we may safely conclude that it should be less than 1 km/s. 
%
\begin{table}[h!]
\centering
\caption{Overview of the membership quality in the V1062 Sco moving group. In Column 1 we show the different membership qualities as explained in the Text, column 2 refers to TGAS stars from Table~\ref{T}, column 3 to HSOY stars with TGAS paprallaxes (Table~\ref{TH}),   column 4 to HSOY stars  (Table~\ref{H}). Finally, in column 5 we give the total number of stars with the respective membership quality. }
\begin{tabular}{|c|r|r|r|r|}
\hline
  \multicolumn{1}{|c|}{Menbership} &
  \multicolumn{1}{c|}{Tab.~\ref{T}} &
  \multicolumn{1}{c|}{Tab.~\ref{TH}}&
  \multicolumn{1}{c|}{Tab.~\ref{H}} &
  \multicolumn{1}{c|}{all} \\
\hline
  m(6d)            & 11  & 1 & - & 12\\
  m(5d) [$\varpi_{trig}$/no $RV$] & 22  & 5 & - & 27\\
  m(5d) [$RV$/no $\varpi_{trig}$] &  - & - & 4 & 4\\
  m(4d)            & -  & - & 20 & 20 \\
  nm            & 2  & 3 & 8 & 13\\
\hline         
  all           & 35  & 9 & 32 & 76\\
\hline\end{tabular}
\label{mem_stat}
\end{table}
\subsection{Colour-magnitude diagrams}
\label{CMD}
\begin{figure}[h!]
   \centering
      \includegraphics[width=0.45\textwidth]{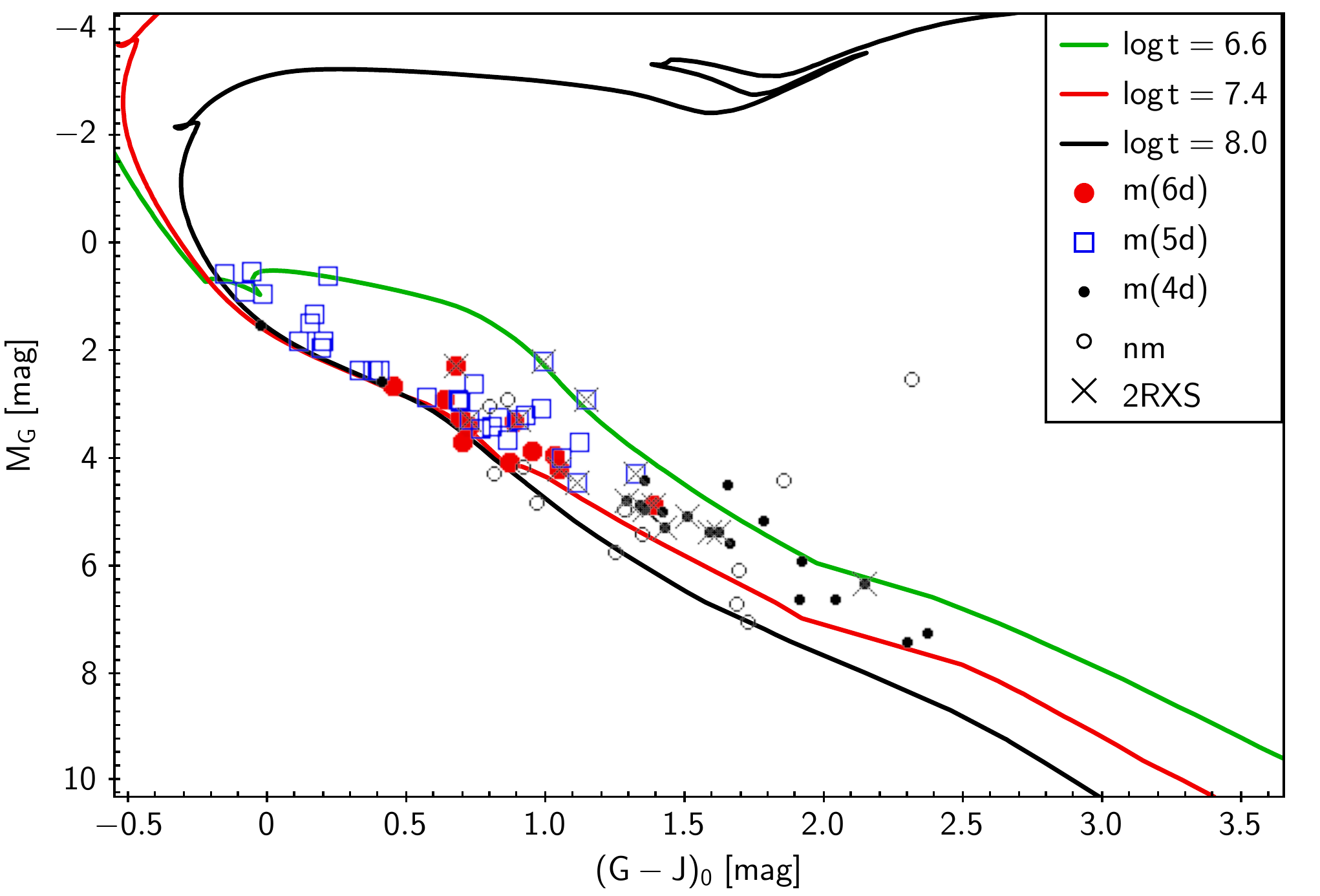}
      \caption{The $M_G$ vs. ${(G-J)}_0$ colour-magnitude diagram of stars in the  V1062 Sco moving group. The symbols for the  members of different membership quality are shown in the legend. \BGS{Crosses mark stars that were observed by the ROSAT satellite (2RXS, see text).}
      Also plotted are three Padova-2.8 Solar metallicity isochrones for log t = 6.6, 7.4 and 8.0.}
         \label{G-J}
   \end{figure}

\begin{figure}[h!]
   \centering
   \includegraphics[width=0.45\textwidth]{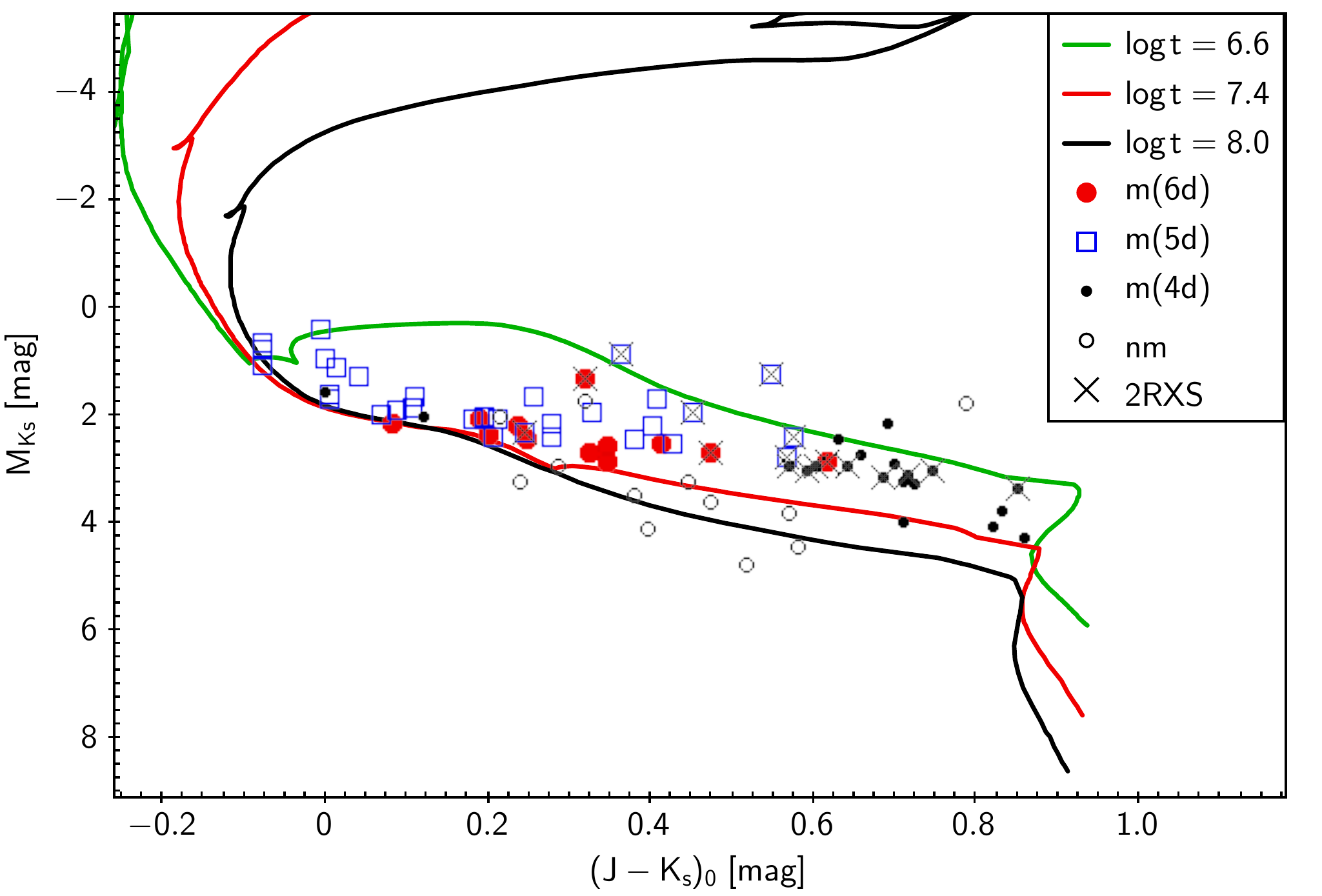}
      \caption{The $M_{K_{S}}$ vs. ${(J-K_{S})}_0$ colour-magnitude diagram of stars in the  V1062 Sco moving group. The symbols for the members of different membership quality are shown in the legend. \BGS{Crosses mark stars that were observed by the ROSAT satellite (2RXS, see text).} Also plotted are three Padova-2.8 Solar metallicity isochrones for log t = 6.6, 7.4 and 8.0.}
         \label{JJK}
   \end{figure}

%
All stars of the V1062 Sco MG as selected above have 2MASS  \citep{cat2MASS} photometry and Gaia DR1  \citep{2016A&A...595A...2G}  $G$-magnitudes.
We present them in the colour-magnitude diagrams (CMDs) $M_G$ vs. $G-J$ in Fig.~\ref{G-J} and $M_{K_{S}}$ vs. $J-K_{S}$ in Fig.~\ref{JJK}. 
The magnitudes on the ordinate are absolute magnitudes, and the colours on the abscissae are intrinsic colours in both diagrams. For the distance moduli we used the $\varpi_{kin}$, first because they are available for all the stars in the 3 tables, and second because, within the TGAS precision budget, they are formally more precise than the $\varpi_{trig}$. In any case the mean difference between the distance moduli from $\varpi_{trig}$ and $\varpi_{kin}$ \BG{is $-0.01$~mag} with a standard deviation of 0.11~mag, so hardly visible on the plots.  
\BGS{We also do not show error bars for the absolute magnitudes. For TGAS stars the median of the uncertainties in distance modulus is 0.1 mag based on kinematic, 0.17 mag on trigonometric parallaxes. For HSOY stars the median is 0.26 mag.  All these quantities give error bars smaller than the size of the symbols.}
The reddening $E(B-V)$ was determined from photometric data on TGAS members since their actual distances are sufficiently well known from the trigonometric parallaxes. We obtained a value of 0.065 mag for $E(B-V)$ from the star's loci in the three CMDs, i.e., $M_G$ vs. $G-J$, $M_{K_{S}}$ vs. $J-K_{S}$ and $M_V$ vs. $B-V$. 
%
Padova-2.8 Solar metallicity isochrones \citep{2012MNRAS.427..127B} for $\log t$ = 6.6, 7.4 and 8.0 are also shown in  Figs.~\ref{G-J} and \ref{JJK}. From the distribution of the stars in both CMDs we estimate that the masses of our members range between 2.6 $\rm{M}_\odot$ to 0.7 $\rm{M}_\odot$.

It is conspicuous in both CMDs that the stars do not follow a unique isochrone, but are mainly located between isochrones of 4 Myrs and 25 Myr, especially those with spectral types later than about F0 ($J-K_{S} > 0.2$). This spread cannot be compensated by inaccuracy of photometric data: a typical $rms$-error is about 0.025~mag for $J$, $K_{S}$, and 0.01~mag for $G$-magnitudes. Also, binarity can alleviate the age discrepancy, but not completely remove it.  A possible explanation can be that the stars have formed over a period of 20 Myrs, but are still moving together with the same space velocity. We \BG{remind} the reader that the V1062 Sco MG is embedded in the Sco-Cen OB association, where many small star forming regions are present (see the discussion in the next \BG{section}), but, to our knowledge, no one is found which is so nicely co-moving and well populated as the V1062 Sco MG.
%
   
We also tried to recover our stars in other sky surveys to get further indication on their youth.  We found that eight TGAS stars (3 m(6d), 5 m(5d)) and nine HSOY14 stars (1 m(5d), 8 m(4d))
have ROSAT-2RXS \citep{2016A&A...588A.103B} counterparts,
but none of them is measured by Galex \BGS{\citep{2007ApJS..173..682M}} as our moving group is outside the area that Galex observed.

As we mentioned above, the convergent point method predicts the parallaxes of the kinematic members. In contrast to TGAS candidates, 
we cannot yet verify the predictions for HSOY stars by measurements. Therefore, some few HSOY kinematic candidates may happen to be 
 non-members of the V1062 Sco MG. We can decrease a possible contamination by checking the consistence of predicted parallaxes and photometric distances of m(4d) candidates. To be on a safe side, we marked a candidate as a non-member when it is located outside the area bordered by the
isochrones $\log t$ = 6.6, 7.4 in at least one of the CMDs ($M_G$ vs. $G-J$,$M_{K_{S}}$  vs. $J-K_{S}$). Five of eight rejected candidates  indeed seemed to be back- or foreground stars, whereas three others were probably main sequence stars at the distance of the V1062 Sco MG. In this case they may be co-moving field stars older than 100 Myrs that are likely not connected to the young moving group V1062 Sco, i.e. "contaminants". 
\subsection{Estimate of the contamination}

To estimate the number of field star contaminants that may enter our sample, we ran a Besan\c{c}on simulation of the area around the cluster centre, obtaining a mock catalogue of field objects with 6D parameters. 
The Besan\c{c}on simulation was performed over 40 square degrees with default parameters. The area was centered at $l=343\fdg7$, $b=+4\fdg3$ and a constant extinction of $A_V=1.2$\,mag/kpc was added. This agrees with  $A_V=0.2$\,mag at 175\,pc, i.e. at the mean distance of the V1062 Sco MG where we estimated $E(B-V)$ to be 0.065~mag. 
We spread the mock positions over the whole area \BGS{of 40 square degrees} assuming a \BGS{uniform distribution on the sky, the line-of-sight distance being determined by the Besan\c{c}on simulator}.
The simulated cartesian parameters were converted into the observables, and we added Gaussian noise of 0.3\,mas/yr~(RMS) and 0.3\,mas~(RMS) on each proper motion component and the parallax, respectively.
We also converted the Besan\c{c}on Cousins photometry into the Gaia G system using the \citet{Jordi10} colour transformation.
Finally, to improve the statistics, we repeated the simulation ten times.

\begin{figure}[h!]
   \centering
   \includegraphics[width=0.45\textwidth]{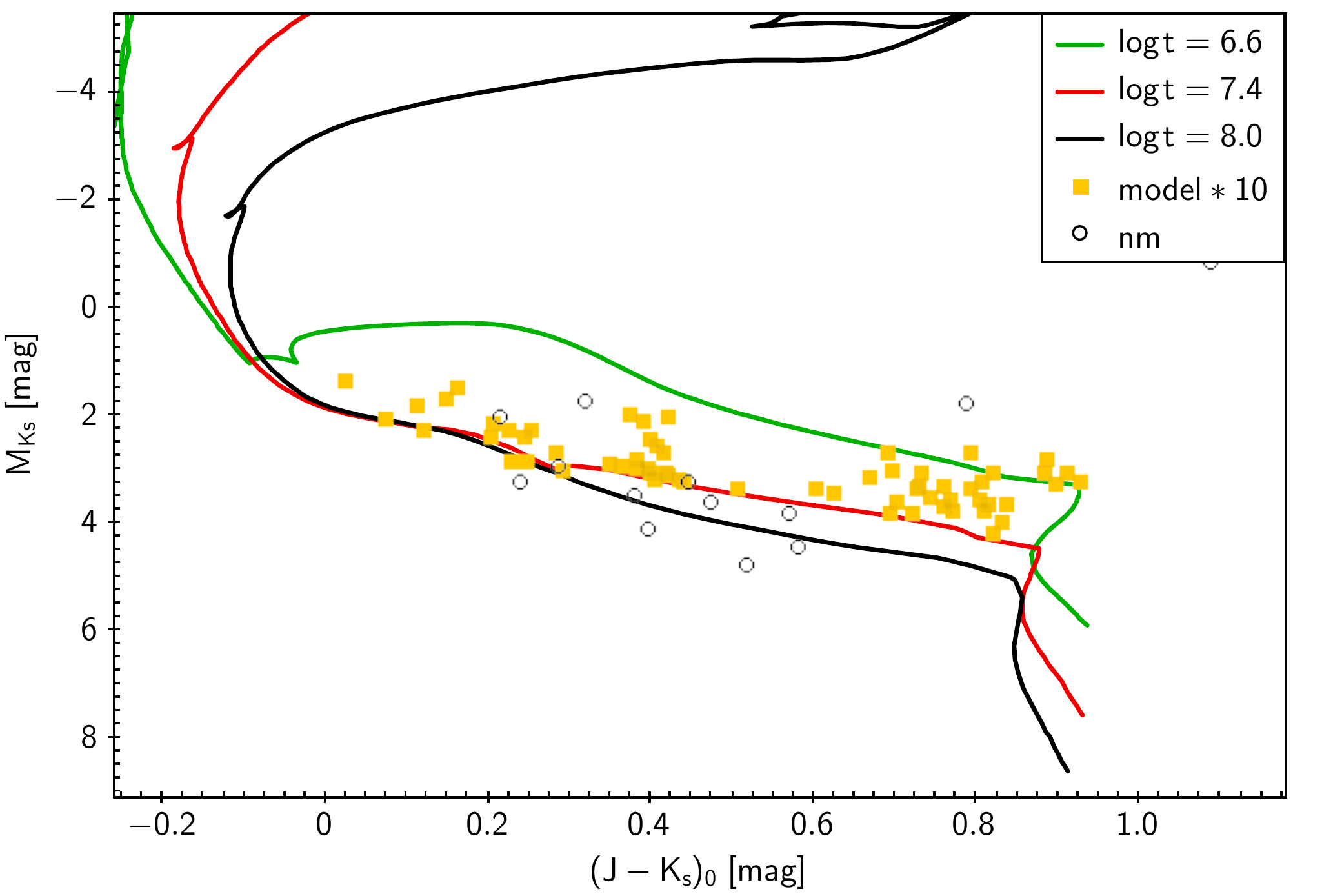}
      \caption{The $M_{K_{S}}$ vs. ${(J-K_{S})}_0$ colour-magnitude diagram of stars in the area of the V1062 Sco MG obtained from a simulation based on the Besan\c{c}on model (10-times enhanced). Again three Padova-2.8 Solar metallicity isochrones for log t = 6.6, 7.4 and 8.0 are also given. Also plotted are the 13 non-members from Figures~\ref{G-J} and \ref{JJK} and Table~\ref{mem_stat}.}
         \label{Simul}
   \end{figure}

As for the real TGAS and HSOY14 stars, we ran the CP method around the phase-space coordinates (\ref{finalxyz}) for this ten times enlarged simulation catalog. In Fig.\,\ref{Simul} we show the simulated stars that came out from the CP method as kinematic candidates i.e., m(4d) members. In total, there are 64 stars in the area between the isochrones for $\log t$ = 6.6 and $\log t$ = 7.4. So, in fact, we expect a number of 7 (6.4) stars from the Besan\c{c}on model to fall into the CMD area occupied by the stars from the V1062 Sco MG.
In the following we give a worst case (maximum) estimate of the contamination: From our 68 kinematic candidates from Table~\ref{mem_stat} in the CMD area we had to discard already 5 as non-members based on discordant parallaxes or radial velocities. So, the maximum contamination among the remaining 63 is two, or 3\%. In reality the contamination is much smaller for the m(5d) or m(6d) members. For instance, if we take into account the simulated parallaxes of the 6.4 contaminants, we find only 0.7 which should fall into the CMD area in Fig.\,\ref{Simul}.

\subsection{Individual stars}
\label{ind_stars}
The star V1062 Sco (star 7 of our sample) is a  variable of $\alpha_2$ CVn type. Its TGAS parallax makes it an m(5d) member in our terminology. In the literature we found a radial velocity measurement of $13.00 \pm 5.80 $ km/s from \citet{1996A&AS..118..231L}. This radial velocity is almost 2-$\sigma$ off the predicted radial velocity. However, in case of a rotationally variable stars  it is hard to extract its line-of-sight velocity. Therefore we keep this star as an m(5d) member. 


Star 5 (HD 149551) and star 18 (HD 150372) are found in \citet{2002AJ....124.1670M}. Depending on different pMS isochrones the authors estimate an age between 4 and 11 Myrs and a mass between 1.5 and 1.7 $\rm{M}_\odot$  for star 5, and  an age between 4 and 7 Myrs and a mass between 1.8 and 1.9 $\rm{M}_\odot$ for star 18. Star 9 (\object{HD 149777}) was rated 'active subgiant' by \citet{2002AJ....124.1670M} and hence rejected as a young member in the Sco-Cen association. However we keep it as a m(5d) member because of its measured trigonometric parallax. \BG{The preliminary velocity measurement ($RV = +0.6\pm 7.8\,$km/s) for this very rapidly rotation star ($v_{\rm rot}\approx 200$\,km/s) confirms its membership. Nevertheless the uncertainty of this measurement is rather large.} 

Star 30 of our sample (HD 152407) is mentioned in SIMBAD as a member in the cluster Trumpler~24. It has number 160 in the catalog of $U,B,V$ measurements in Tr~24 by \citet{1984A&AS...57..205H}. According to Heske \& Wendker, TR~24 has a distance of 2300~pc, and they rate HD 152407 as a post-MS member from their photometry and give its spectral type of B5Ib. However, its trigonometric parallax of 5.48 mas clearly rules it out as a member in Tr~24, and sets doubts on its luminosity class Ib.
 
Star 141 of our sample is SZB86 from \citet{2012AJ....144....8S}. They quote a distance of 99 pc, photometrically estimated from an empirical 10 Myr isochrone on an $M_{K_{S}}$ versus $V - K_{S}$ CMD. They give a typical uncertainty of about 30\% for their determination. This star is also found in \citet{2003A&A...403..247F}, in their catalog of X-ray variable stars as No. 844. They determine its spectral type to be  M0Ve and rate it as a flare star. We keep this star as an m(4d) candidate, because of its kinematic distance of 175~pc.

Star 35 (\object{HD 151738}) is a spectroscopic binary with a difference in radial velocity of about 100\,km/s. Because we do not know  the mass ratio, we cannot use the measurement to constrain its membership.

Given the age range in V1062~Sco MG, we expect no detection of protoplanetary \BGS{disks} and very few detections of warm debris disks \citep{Cieza10}. However, using VOSA \citep{Bayo08} we reveal a significant infrared excess in the WISE \citep{2012yCat.2311....0C} bands for six targets: star 107 in W1 (3.4$\mu$m) redwards, stars 22, 124, and 143 in W3 (12$\mu$m) and W4 (22$\mu$m), and stars 26 and 59 in W4 only.

\BGS{
\begin{itemize}
\item Visual inspection of the relevant WISE images showed that stars 22 (\object{HD 150532}), 124 (\object{2MASS 16375854-3933221}) and 143 are associated with nebulosity both in W3 and W4 bands. Such complex backgrounds can lead to spurious detections \citep[e.g.][]{2014MNRAS.437..391C}. Actually we found no sources at the position of the stars in the W4 band images. The W3 band excess detections should also be taken carefully given the nebulosity.

\item  Star 107 (\object{2MASS 16425498-4233057}) was observed with the {\sl Spitzer Space Telescope} using the IRAC instrument in the 3.6, 4.5, 5.8, and 8.0~$\mu$m bands.  These photometric data were taken from  
the Spitzer Enhanced Imaging Products catalog (SEIP\footnote{http://irsa.ipac.caltech.edu/data/SPITZER/Enhanced/SEIP/docs/seip\_explanatory\_supplement\_v3.pdf}). In contrast to WISE observations, the analysis of IRAC data shows no evidence for excess between 3.6 and 8.0~$\mu$m. The reasons of these discrepancies are not yet clear.

\item The IR excess of star 26 (\object{HD 151109}) was also detected by \citet{Rizzu12} and \citet{Chen12}. Based on photometric and spectroscopic observations with Spitzer, \citet{Chen14} fitted the observed spectral shape of the excess using two blackbody components with temperatures
of 237K and 91K. This two-component debris disk has a total fractional luminosity of $1.1\times10^{-4}$. 

\item The W4 band excess detection at star 59 (\object{HD 152369}) also seems to be solid.
\end{itemize}
}

None of those stars shows signs of accretion in our FEROS spectra.
We ran a similar analysis on 221 random TGAS stars located at the same distance. None of those random stars showed consistent infrared excess as significant as our cluster members, giving a good hint that those disk detections may be real.
\section{The V1062 Sco Moving Group and its relation to the Upper Centaurus Lupus association}
The newly detected object is a co-moving group of young stars at the edge of the huge (845 deg$^2$) UCL association. The ages of the V1062 Sco MG members  range from less than 10 to about 25 Myrs. Some of the younger stars were observed by the ROSAT satellite and show X-ray activity. The distance of the group from the Sun is  $175 \pm 7$ pc. Within the UCL complex there are a number of areas where star formation has recently taken place or is still going on. These areas are related to dense molecular clouds in UCL and have been studied by a number of authors in the past decades. However, no evidence has been obtained for a compact moving group around the variable star V1062 Sco.
%
   \begin{figure}[t!] 
   \centering
   \includegraphics[width=0.45\textwidth]{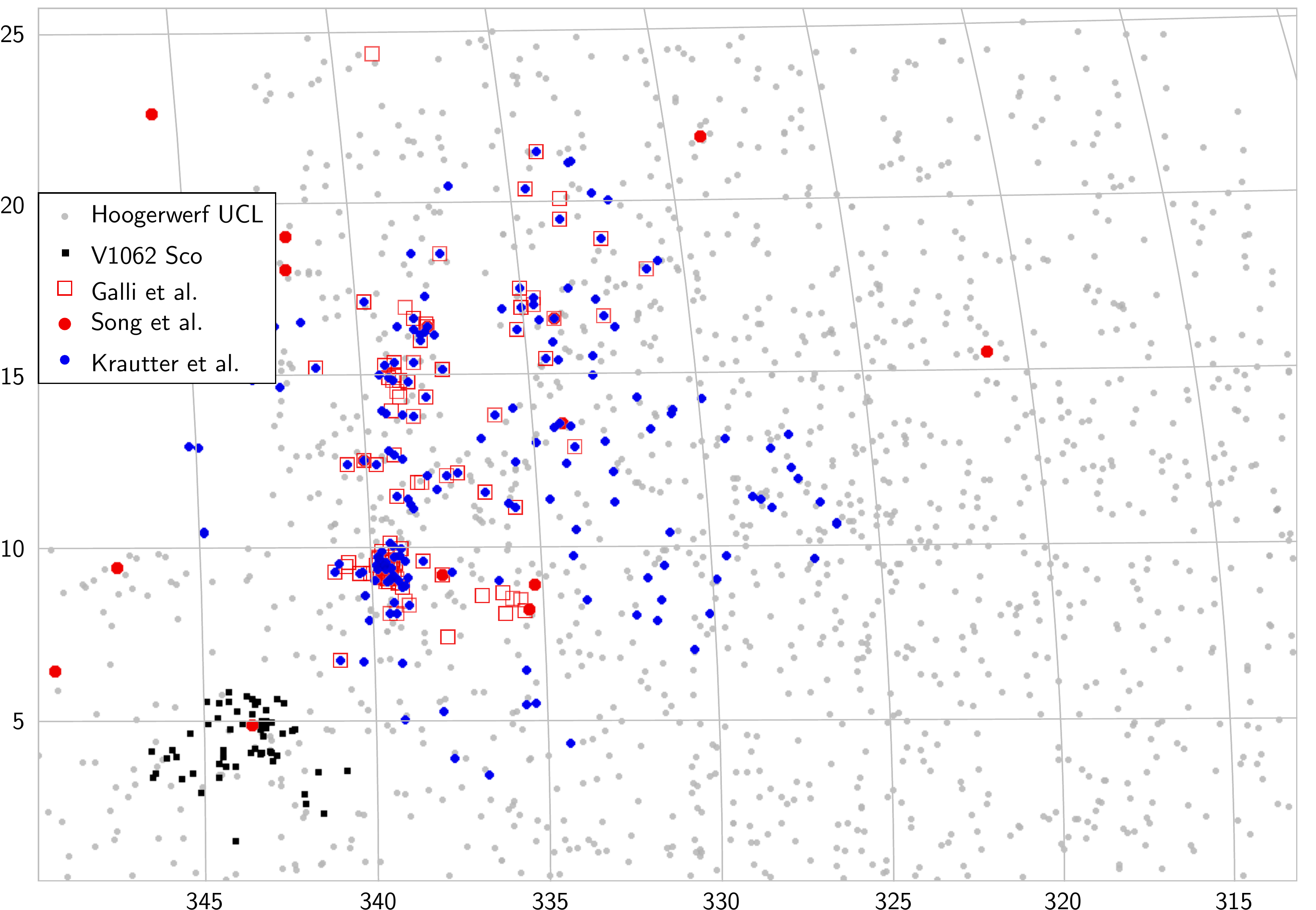}
      \caption{The area of UCL in Galactic coordinates. The extent of UCL is \BGS{outlined} by the distribution of the stars from \citet{2000MNRAS.313...43H}, (grey dots). Our 63 members of the V1062 Sco MG are shown as black squares. Young stars from other authors in the UCL area are shown with the symbols indicated on the map.}
         \label{UCLarea}
   \end{figure}

In his study of the Sco~OB2\verb+_+3 (UCL) area, \citet{2000MNRAS.313...43H} searched for stars in the TRC catalog \citep{1998A&A...335L..65H}, a predecessor of Tycho-2, which had proper motions consistent with the spatial velocity of the Hipparcos members of UCL from  \citet{1999AJ....117..354D}. Using a convergent-point method, with further constraints on the proper motion distribution, magnitude and colour, he found some 1200 candidate stars. Based on Monte-Carlo simulations of the field star population he expected a contamination of between 400 and 800 field stars in this sample. He thus concluded that, among the 1200 stars, a considerable number of  association members should be present which are fainter than the about 150 Hipparcos stars from  \citet{1999AJ....117..354D}. 

\begin{figure}[t]
   \centering
   \includegraphics[width=0.45\textwidth]{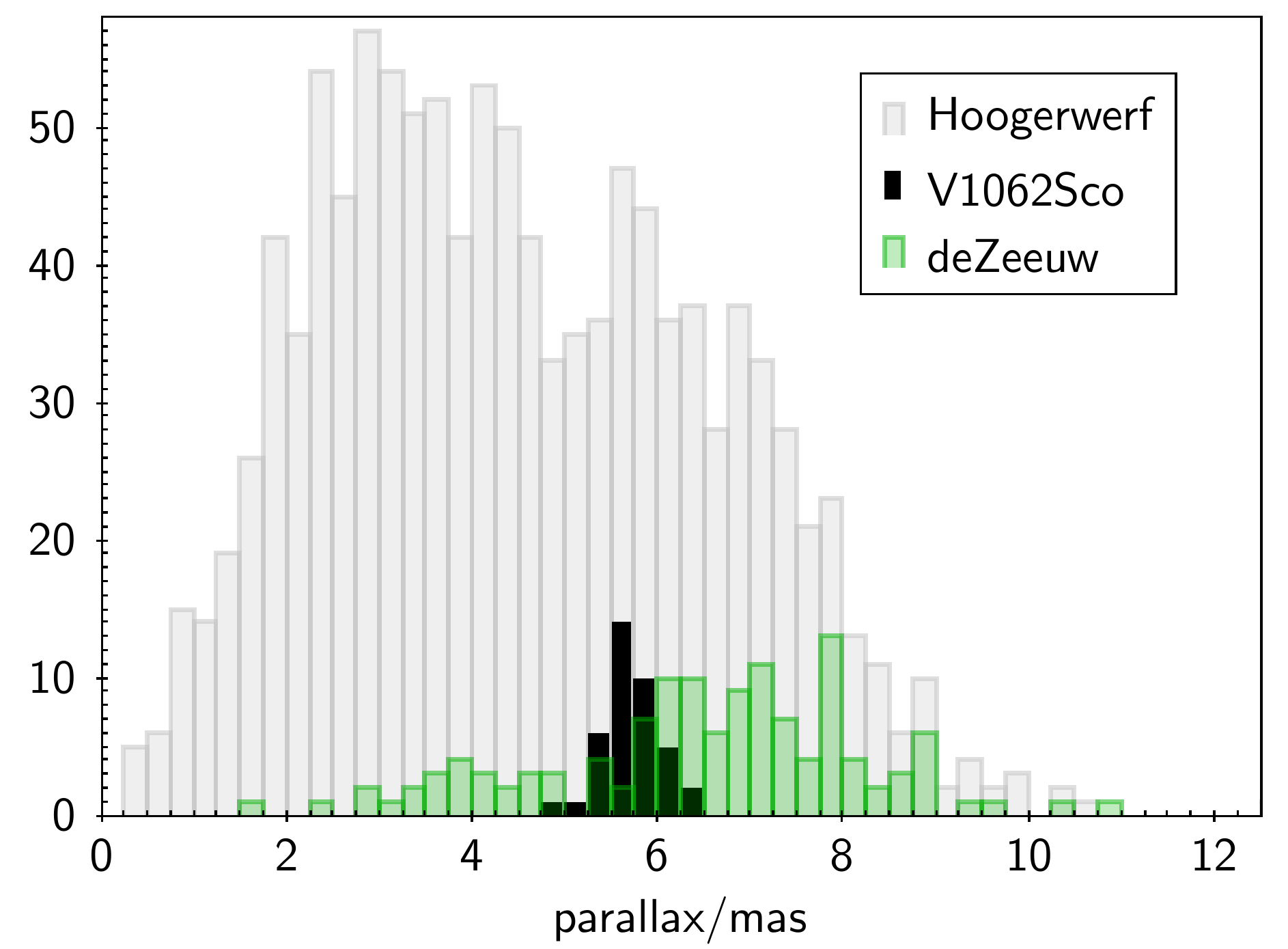}
      \caption{Parallax distribution of stars in the UCL association. Grey bars show the TGAS parallaxes of the stars from \citet{2000MNRAS.313...43H}, green bars the Hipparcos stars from \citet{1999AJ....117..354D}. The black bars display the TGAS parallaxes of the 39 stars from Table~\ref{mem_stat} which are at least m(5d) members.}
         \label{parallaxes}
   \end{figure}

In Fig.~\ref{UCLarea}, we show Hoogerwerf's stars as small grey dots in the background. These stars outline the extent of the UCL area. From the 35 stars in Table~\ref{T} 27 are listed in \citet{2000MNRAS.313...43H}, but only six are also contained in \citet{1999AJ....117..354D}. Also six stars in Table~\ref{TH} and four in Table~\ref{H} are included in \citet{2000MNRAS.313...43H}. We crossmatched the UCL association candidates with TGAS, and show in Fig.~\ref{parallaxes} their distribution versus the trigonometric parallaxes. They demonstrate a broad distribution with parallaxes between 1 mas and 10 mas. There is \BG{indicative} of a bi-modal distribution with one part between 1 and 5 mas, probably the background field-star contaminants mentioned above, and between 5 and 8 mas of the UCL association proper. Also in Fig.~\ref{parallaxes} we show the distribution of the Hipparcos stars (but with TGAS parallaxes) from \citet{1999AJ....117..354D}. Their parallaxes are concentrated between  
 6~mas and 8~mas, showing a lower number of background contaminants. On top the parallax distributions in the UCL area in  Fig.~\ref{parallaxes},  we plot the distribution of TGAS parallaxes of our 39 stars from Table~\ref{mem_stat} which are at least m{5d} members. 
The V1062 Sco moving group members show a narrow parallax distribution clearly indicating a sub-condensation at the periphery of the UCL complex.
%

 
Based on X-ray data from the  ROSAT All-Sky-Survey \citet{1997A&AS..123..329K} discovered 136 new TTauri stars in the Lupus SFR (their tables 5, 6, and 11).  
The positions of these stars in UCL are also shown in  Fig.~\ref{UCLarea}. For 31 TTauri stars in \citet{1997A&AS..123..329K} we identified TGAS counterparts. In comparison to our V1062 Sco MV members, they show a much wider parallax spread, between 5 and 11 mas with a peak at 7.5 mas (ca. 133 pc).

\citet{2012AJ....144....8S} have spectroscopically identified 100~\mbox{G-}, \mbox{K-}, and M-type members of the Scorpius-Centaurus complex. Sixteen of them are found in the UCL complex and  are also plotted in  Fig.~\ref{UCLarea}. As already mentioned, their star SZB86 is an m(4d) kinematic member of our group (star 141). \citet{2012AJ....144....8S} derived ages of about 10 Myr for stars in the Upper Centaurus Lupus region. 

\citet{2013A&A...558A..77G} identified a comoving group with 109 pre-main sequence stars and candidates in the 
Lupus SFR. Assuming that all stars share the same space motion, they derived individual parallaxes for stars with known radial velocity and tentative parallaxes for the remaining group members.  In their Fig.~3 they show the location of their UCL stars overplotted to the $^{12}$CO intensity map from \citet{2001PASJ...53.1081T}. \citet{2013A&A...558A..77G} studied four sub-groups called Lupus 1 to Lupus 4, all associated with strong $^{12}$CO emission. The average kinematic distances of these sub-groups range from 140 to 200~pc. Lupus 3, the most populated group at $(l,b) = (339\fdg5,+9\fdg5)$ in Fig.~\ref{UCLarea} has an average kinematic distance of 180~pc, but shows a considerable depth, i.e. the stars in Lupus 3 are located  between 160 and 260~pc. Also the proper motions of this sample scatter between -5 and -35~mas/yr in both components.   In a later paper, \citet{2015A&A...580A..26G} derived  
masses and ages of the TTS population in this SFR. The ages of their stars peak between 6 and 7 Myrs, and are, on average, younger than ours.  

Unlike the other star forming regions in Lupus which are embedded in molecular clouds and show considerable scatter in their proper motions, the V1062 Sco MG is concentrated in a distance range between 165 and 185~pc, the scatter of its proper motions is comparable to the uncertainties of the proper motions in TGAS, resp. HSOY, and, finally, it is not embedded in a molecular cloud on the \citet{2001PASJ...53.1081T} $^{12}$CO map. In Section~\ref{CMD} we found a moderate value for $E(B-V)$, i.e. an $A_V$ of about 0.2 mag in the direction of our group. 

We additionally studied the interstellar-medium gas absorption in the NaI lines at 589.0 and 589.6\,nm
\citep{Pascu15}. Many of our targets show narrow-line absorption on top of the broad stellar absorption. These lines have a mean barycentric radial velocity of $+0.5$\,km/s (in the LSR: +6\,km/s), with an $rms$-spread of 2\,km/s. Over most of the cluster, the stellar and gas RVs are very similar.
While the RV spread is random on the sky, the depth of the absorption line is noticeably strong (mean of 71\%) west of a great circle passing through $l,b=(343\degr,+4\degr)$, with a orientation of -10$\degr$ (N to E), and small (mean of 31\%) east of that line.
This drop can be seen on the H$_\alpha$ map of this velocity \citep{HI4PI16}. 
With these data, it is not possible to place the gas \BGS{relative} to the
cluster and claim that the cluster still holds some of its original gas.
We can however describe HI emission seen at other RV ($>+10$\,km/s or $<-2$\,km/s in the LSR) by \BGS{the} \citet{HI4PI16} as background gas.
%
\section{Summary}
The Upper Centaurus Lupus (UCL) subgroup of the Sco-Cen OB-association contains a large number of some 130 B and A stars
at a mean distance of 140 pc from the Sun. Also \BGS{it comprises} a few subgroups, embedded in molecular clouds, where recent stars formation is going on. The new group around the variable star V1062 Sco is situated outside \BGS{of} these molecular clouds at a mean distance of $175 \pm 7$ pc from the Sun and contains 
63 stars within a radius of 10~pc around the center. \BGS{The stars in our moving group are not coeval, but show an age spread between 4 and 25 Myr for which we have no explanation}. This group is strongly co-moving, the one-dimensional velocity dispersion being less than 1~km/s.
The 33 members
from the TGAS catalogue are selected such that the difference between their kinematic and trigonometric parallaxes is smaller than the 1-$\sigma$ error of this difference.
For a subset of 12 stars, we could confirm a full 3-D coincidence of their space motion with the mean space motion of the
group. These definitely form a moving group, or possibly an open cluster. For another subset of 27 stars the kinematically predicted
parallaxes are confirmed by the trigonometric parallaxes from TGAS, they fulfil this necessary condition of being cluster members. Also four stars from HSOY have radial velocities confirmed. A last subset of 20 stars have predicted parallaxes consistent with the condition of co-movement. 
In total there are 63 stars belonging to this V1062 Sco moving group. A possible contamination by field stars is about 3\% estimated on the basis of the Besan\c{c}on model of our Galaxy.  In summary, we have clearly shown that the V1062 Sco moving group is a new compact, young moving group within 200~pc from the Sun. The masses of our 63 members range from 2.6 $\rm{M}_\odot$ to 0.7 $\rm{M}_\odot$. With Gaia data release 2, announced for April 2018 we expect to reveal new and fainter members of the group. 

%
%

\begin{acknowledgements}
\BG{BG thanks Andres Jordan and Rafael Brahm for useful discussions and their help in reducing the FEROS data, as well as Caroline Bot and Jeroen Bouwman for their insights. The authors thank Ilaria Pascucci for her precious suggestions regarding IR excess and gas absorption.}
This study was supported by Sonderforschungsbereich SFB 881 "The Milky Way
System" (subprojects B5 and B7) of the German Research Foundation (DFG). 
This research has made use of the SIMBAD database and of the VizieR catalogue access tool, operated at CDS, Strasbourg, France.
This work has made use of data from the European Space Agency (ESA)
mission {\it Gaia} (\url{https://www.cosmos.esa.int/gaia}), processed by
the {\it Gaia} Data Processing and Analysis Consortium (DPAC,
\url{https://www.cosmos.esa.int/web/gaia/dpac/consortium}). Funding
for the DPAC has been provided by national institutions, in particular
the institutions participating in the {\it Gaia} Multilateral Agreement. This publication makes use of VOSA, developed under the Spanish Virtual Observatory project supported from the Spanish MICINN through grant AyA2011-24052.
This project has been supported by the Hungarian NKFI Grants K-115709 and K-119517 of the Hungarian National Research, Development and Innovation Office. AD was supported by the \'UNKP-17-4 New National Excellence Program of the Ministry of Human Capacities. AD would like to thank the City of Szombathely for support under Agreement No. 67.177-21/2016.
\BGS{We would like to thank the anonymous referee for her/his very constructive and helpful comments.}
\end{acknowledgements}




%
%
\bibliographystyle{aa}
\bibliography{mybib}
\begin{appendix}
\section{Tables}
\label{alltables}
\begin{footnotesize}

\begin{sidewaystable*}
\caption{
\BGF{TGAS stars in the V1062 Sco moving group.}
\newline
The columns of the table give: Col. 1 the internal star number in the V1062Sco MG,
col. 2 the main identifier of the star in SIMBAD.
Cols. 3 and 4 the right ascension and declination,
cols. 5 and 6 the proper motions in right ascension and declination in the TGAS catalog.
Col. 7 the kinematic parallax and its mean error from the CP method,
col. 8 the trigonometric parallax and its mean error from TGAS.
Cols. 9 to 11 the $J, H, K_S$ magnitudes from 2MASS,
col. 12 the mean $G$-magnitude from Gaia DR1,
col. 13 gives the membership information (and comments), 
cols. 14 to 16 the cross-match with \citet{1999AJ....117..354D}, \citet{2000MNRAS.313...43H} and  ROSAT-2RXS \citep{2016A&A...588A.103B}.
}
\label{T}
\begin{tabular}{|r|l|c|c|c|c|r@{$\pm$}l|r@{$\pm$}l|c|c|c|c|c|c|c|c|}
\hline
nr  & Name & RA (J2000) & DEC (J2000) & $\mu_\alpha$ & $\mu_\delta$ & \multicolumn{2}{c|}{$\varpi_{\rm kin.}$} & \multicolumn{2}{c|}{$\varpi_{\rm TGAS}$} & $J$ & $H$ & $K_{\rm s}$ & $G$ & Membership & Z & H & X \\
  &  & (deg) & (deg) & (mas/y) & (mas/y) & \multicolumn{2}{c|}{(mas)} & \multicolumn{2}{c|}{(mas)} & (mag) & (mag) & (mag) & (mag) &  &  &  &  \\
\hline
  1  & \object{HD 149354     } & 248.980605 & -40.333009 & -14.11 & -20.41 & 5.69 & 0.29 & 5.53 & 0.44 & 8.542 & 8.266 & 8.178 & 9.589 & m(5d)            & - & H & - \\
  2  & \object{HD 149425     } & 249.119405 & -40.303120 & -11.92 & -20.84 & 5.52 & 0.01 & 5.71 & 0.51 & 6.744 & 6.781 & 6.714 & 7.075 & m(5d)            & Z & H & - \\
  3  & \object{HD 149467     } & 249.182939 & -39.332763 & -12.79 & -20.90 & 5.62 & 0.32 & 5.50 & 0.25 & 8.738 & 8.517 & 8.426 & 9.685 & m(5d)            & - & H & - \\
  4  & \object{HD 149533     } & 249.260954 & -38.676307 & -12.53 & -20.98 & 5.61 & 0.01 & 6.30 & 0.42 & 7.268 & 7.273 & 7.233 & 7.372 & m(5d)            & - & H & - \\
  5  & \object{HD 149551     } & 249.303557 & -39.010658 & -11.62 & -21.03 & 5.51 & 0.02 & 5.76 & 0.39 & 8.453 & 8.103 & 8.010 & 9.553 & m(5d)            & Z & H & - \\
  6  & \object{TYC 7858-585-1} & 249.425372 & -40.171494 & -11.55 & -21.04 & 5.51 & 0.50 & 6.01 & 0.92 & 9.273 & 8.943 & 8.826 & 10.426 & m(6d)            & - & - & - \\
  7  & \object{V* V1062 Sco  } & 249.628854 & -39.152419 & -12.58 & -21.21 & 5.66 & 0.01 & 5.82 & 0.44 & 6.883 & 6.927 & 6.926 & 6.941 & m(5d)            & - & H & - \\
  8  & \object{HD 149726     } & 249.630042 & -41.635380 & -13.67 & -21.19 & 5.80 & 0.03 & 6.24 & 0.28 & 8.488 & 8.325 & 8.270 & 9.302 & m(5d)            & - & H & - \\
  9  & \object{HD 149777     } & 249.660225 & -39.551078 & -11.83 & -21.02 & 5.54 & 0.02 & 5.69 & 0.39 & 7.561 & 7.280 & 7.161 & 8.670 & m(5d)            & - & H & X \\
 10  & \object{HD 149793     } & 249.681434 & -39.679564 & -13.07 & -22.60 & 6.00 & 0.35 & 5.55 & 0.22 & 8.618 & 8.420 & 8.345 & 9.625 & m(6d)            & - & H & - \\
 11  & \object{HD 149938     } & 249.888793 & -39.609281 & -12.16 & -21.14 & 5.60 & 0.12 & 5.49 & 0.36 & 7.939 & 7.891 & 7.898 & 8.260 & m(5d)            & - & H & - \\
 12  & \object{HD 150004     } & 249.991238 & -39.789033 & -12.11 & -20.68 & 5.50 & 0.09 & 5.54 & 0.34 & 7.499 & 7.494 & 7.451 & 7.785 & m(5d)            & - & H & - \\
 13  & \object{HD 150107     } & 250.137078 & -38.226872 & -11.95 & -22.55 & 5.85 & 0.27 & 5.63 & 0.31 & 8.785 & 8.603 & 8.548 & 9.598 & m(6d)            & - & H & - \\
 14  & \object{HD 150092     } & 250.217787 & -40.433294 & -13.04 & -22.30 & 5.94 & 0.60 & 6.19 & 0.89 & 9.099 & 8.824 & 8.716 & 10.169 & m(6d)            & - & H & - \\
 15  & \object{HD 150196     } & 250.284884 & -40.225852 & -12.56 & -21.31 & 5.68 & 0.29 & 5.59 & 0.26 & 8.970 & 8.737 & 8.686 & 9.809 & m(6d)            & - & H & - \\
 16  & \object{HD 150287     } & 250.466614 & -40.466324 & -11.76 & -21.41 & 5.61 & 0.02 & 5.79 & 0.28 & 8.581 & 8.388 & 8.357 & 9.338 & m(6d)            & Z & H & - \\
 17  & \object{TYC 7871-694-1} & 250.500567 & -40.049478 & -11.71 & -21.39 & 5.60 & 0.63 & 5.67 & 0.29 & 9.812 & 9.312 & 9.159 & 11.318 & m(6d)            & - & - & X \\
 18  & \object{HD 150372     } & 250.599928 & -40.058326 & -10.59 & -22.63 & 5.72 & 0.27 & 5.30 & 0.48 & 8.047 & 7.614 & 7.464 & 9.310 & m(5d)            & - & - & X \\
 19  & \object{HD 321892     } & 250.613520 & -38.460159 & -12.91 & -21.52 & 5.75 & 0.42 & 5.84 & 0.23 & 8.642 & 8.245 & 8.156 & 9.660 & m(5d)            & - & H & X \\
 20  & \object{HD 150398     } & 250.627049 & -38.930180 & -11.28 & -21.87 & 5.64 & 0.23 & 5.80 & 0.25 & 8.522 & 8.433 & 8.404 & 9.092 & m(6d)            & - & H & - \\
 21  & \object{HD 325907     } & 250.631351 & -41.759400 & -12.43 & -22.55 & 5.93 & 0.45 & 5.53 & 0.25 & 9.416 & 9.090 & 9.034 & 10.401 & m(6d)            & - & - & - \\
 22  & \object{HD 150532     } & 250.872192 & -41.986154 & -9.54 & -24.10 & 5.92 & 0.31 & 5.70 & 0.24 & 8.802 & 8.679 & 8.561 & 9.725 & m(5d)            & - & H & - \\
 23  & \object{HD 150989     } & 251.554962 & -39.745820 & -11.99 & -22.91 & 5.93 & 0.17 & 5.91 & 0.30 & 8.182 & 8.128 & 8.040 & 8.680 & m(5d)            & - & H & - \\
 24  & \object{HD 150988     } & 251.570570 & -39.272726 & -11.43 & -22.03 & 5.69 & 0.10 & 5.67 & 0.35 & 7.987 & 7.971 & 7.946 & 8.215 & m(5d)            & - & H & - \\
 25  & \object{HD 321958     } & 251.667832 & -38.147518 & -11.83 & -21.74 & 5.67 & 0.62 & 5.91 & 0.25 & 9.452 & 9.030 & 8.942 & 10.621 & m(6d)            & - & - & X \\
 26  & \object{HD 151109     } & 251.756929 & -39.533964 & -11.59 & -21.97 & 5.70 & 0.01 & 4.96 & 0.91 & 6.998 & 7.038 & 7.039 & 6.963 & m(5d)            & Z & H & - \\
 27  & \object{HD 151681     } & 252.602617 & -38.048545 & -12.45 & -22.22 & 5.83 & 0.13 & 5.82 & 0.30 & 7.982 & 7.968 & 7.838 & 8.295 & m(5d)            & - & H & - \\
 28  & \object{HD 151726     } & 252.689598 & -38.256513 & -11.79 & -22.70 & 5.86 & 0.01 & 5.69 & 0.39 & 7.218 & 7.261 & 7.261 & 7.257 & m(5d)            & Z & - & - \\
 29  & \object{HD 152041     } & 253.173750 & -38.760443 & -11.32 & -22.02 & 5.67 & 0.02 & 5.38 & 0.37 & 7.913 & 7.727 & 7.558 & 8.706 & m(6d)            & Z & H & X \\
 30  & \object{HD 152407     } & 253.743519 & -41.092913 & -12.90 & -21.79 & 5.80 & 0.20 & 5.48 & 0.37 & 8.213 & 8.155 & 8.089 & 8.735 & m(5d)            & - & H & - \\
 31  & \object{HD 321803     } & 248.548464 & -39.626711 & -9.30 & -22.78 & 5.60 & 1.28 & 6.32 & 0.23 & 9.271 & 8.965 & 8.809 & 10.442 & m(5d)            & - & H & - \\
 32  & \object{HD 149230     } & 248.827830 & -42.239950 & -12.99 & -21.23 & 5.74 & 0.27 & 5.42 & 0.39 & 8.495 & 8.383 & 8.265 & 9.294 & m(5d)            & - & H & - \\
 33  & \object{TYC 7358-330-1} & 249.948782 & -36.609805 & -11.92 & -22.38 & 5.80 & 0.80 & 2.15 & 0.27 & 9.909 & 9.594 & 9.427 & 11.311 & nm (kin)         & - & - & - \\
 34  & \object{HD 325890     } & 250.201816 & -41.271889 & -9.78 & -24.05 & 5.92 & 0.35 & 3.13 & 0.26 & 9.450 & 9.222 & 9.129 & 10.487 & nm (kin), binary & - & - & - \\
 35  & \object{HD 151738     } & 252.742709 & -38.928093 & -11.13 & -21.13 & 5.47 & 0.32 & 5.48 & 0.27 & 8.935 & 8.717 & 8.656 & 9.781 & m(5d), binary    & - & H & X \\

\hline
\end{tabular}
\vspace{0.1truecm}
\end{sidewaystable*}
\end{footnotesize}

\begin{sidewaystable*}
\caption{
\BGF{TGAS stars with HSOY proper motions in the V1062 Sco moving group.}
\newline
The columns of the table give: Col. 1 the internal star number in the V1062Sco MG,
col. 2 the main identifier of the star in SIMBAD.
Cols. 3 and 4 the right ascension and declination,
cols. 5 and 6 the proper motions in right ascension and declination in the HSOY catalog.
Col. 7 the kinematic parallax and its mean error from the CP method,
col. 8 the trigonometric parallax and its mean error from TGAS.
Cols. 9 to 11 the $J, H, K_S$ magnitudes from 2MASS,
col. 12 the mean $G$-magnitude from Gaia DR1,
col. 13 gives the membership information (and comments),
cols. 14 and 15 the  cross-match with \citet{2000MNRAS.313...43H} and  ROSAT-2RXS \citep{2016A&A...588A.103B}.}
\label{TH}
\begin{tabular}{|r|l|c|c|c|c|r@{$\pm$}l|r@{$\pm$}l|c|c|c|c|c|c|c|c|c|}
\hline
nr  & Name & RA (J2000) & DEC (J2000) & $\mu_\alpha$ & $\mu_\delta$ & \multicolumn{2}{c|}{$\varpi_{\rm kin.}$} & \multicolumn{2}{c|}{$\varpi_{\rm TGAS}$} & $J$ & $H$ & $K_{\rm s}$ & $G$ & Membership & H & X \\
  &  & (deg) & (deg) & (mas/y) & (mas/y) & \multicolumn{2}{c|}{(mas)} & \multicolumn{2}{c|}{(mas)} & (mag) & (mag) & (mag) & (mag) &  &  &  \\
\hline
 51  & \object{HD 149638                                      } & 249.429970 & -39.708627 & -13.21 & -20.97 & 5.69 & 0.23 & 5.71 & 0.22 & 8.862 & 8.539 & 8.425 & 10.097 & m(5d)    & H & - \\
 52  & \object{HD 150165                                      } & 250.234747 & -40.187218 & -11.27 & -22.49 & 5.77 & 0.15 & 5.83 & 0.28 & 8.304 & 8.049 & 7.950 & 9.280 & nm (kin) & H & - \\
 53  & \object{HD 149738                                      } & 249.612316 & -39.651811 & -11.73 & -20.80 & 5.48 & 0.06 & 6.07 & 0.32 & 7.683 & 7.658 & 7.607 & 7.955 & m(5d)    & H & - \\
 54  & \object{HD 322024                                      } & 251.071987 & -40.252976 & -14.11 & -20.20 & 5.63 & 0.25 & 0.71 & 0.87 & 6.533 & 5.687 & 5.410 & 8.967 & nm (kin) & - & - \\
 55  & \object{HD 149336                                      } & 248.895070 & -37.110095 & -11.44 & -22.61 & 5.80 & 0.15 & 6.17 & 0.25 & 8.486 & 8.249 & 8.237 & 9.399 & nm (RV)  & H & - \\
 56  & \object{HD 149984                                      } & 249.946029 & -38.769026 & -13.02 & -21.86 & 5.83 & 0.22 & 5.79 & 0.22 & 9.232 & 8.951 & 8.872 & 10.051 & m(6d)    & H & - \\
 57  & \object{HD 321981                                      } & 251.376202 & -39.161844 & -11.63 & -21.44 & 5.59 & 0.24 & 5.00 & 0.38 & 9.300 & 8.787 & 8.690 & 10.741 & m(5d)    & - & X \\
 58  & \object{HD 151868                                      } & 252.940040 & -38.052476 & -12.32 & -21.31 & 5.63 & 0.18 & 5.90 & 0.22 & 8.593 & 8.427 & 8.345 & 9.281 & m(5d)    & - & - \\
 59  & \object{HD 152369                                      } & 253.642044 & -38.105640 & -12.05 & -22.56 & 5.85 & 0.09 & 6.24 & 0.92 & 8.262 & 8.204 & 8.159 & 8.710 & m(5d)    & H & - \\

\hline
\hline\end{tabular}
\end{sidewaystable*}

\begin{sidewaystable*}
\caption{\BGF{HSOY stars  in the V1062 Sco moving group.}
\newline
The columns of the table give: Col. 1 the internal star number in the V1062Sco MG,
col. 2 the main identifier of the star in SIMBAD, resp. the 2MASS identifier.
Cols. 3 and 4 the right ascension and declination,
cols. 5 and 6 the proper motions right ascension and declination in the HSOY catalog.
Col. 7 the kinematic parallax and its mean error from the CP method.
Cols. 8 to 10 the $J, H, K_S$ magnitudes from 2MASS,
col. 11 the mean $G$-magnitude from Gaia DR1,
cols. 12 and 13 give the membership information and comments,
cols. 14 and 15 the  cross-match with \citet{2000MNRAS.313...43H} and  ROSAT-2RXS \citep{2016A&A...588A.103B}.}
\label{H}
\begin{tabular}{|r|l|c|c|c|c|r@{$\pm$}l|c|c|c|c|c|c|c|c|}
\hline
\hline
nr  & Name & RA (J2000) & DEC (J2000) & $\mu_\alpha$ & $\mu_\delta$ & \multicolumn{2}{c|}{$\varpi_{\rm kin.}$} & $J$ & $H$ & $K_{\rm s}$ & $G$ & Membership & H & X \\
  &  & (deg) & (deg) & (mas/y) & (mas/y) & \multicolumn{2}{c|}{(mas)} & (mag) & (mag) & (mag) & (mag) &  &  &  \\
\hline
101  & \object{16365957-4137069} & 249.248204 & -41.618651 & -12.05 & -21.72 & 5.72 & 0.68 & 11.292 & 10.796 & 10.676 & 13.093 & nm           & - & - \\
103  & \object{16302731-3915237} & 247.613815 & -39.256619 & -11.91 & -21.48 & 5.64 & 0.68 & 11.624 & 11.235 & 11.070 & 13.463 & nm           & - & - \\
105  & \object{16342105-3922198} & 248.587707 & -39.372174 & -12.15 & -22.25 & 5.82 & 0.81 & 10.049 & 9.448 & 9.297 & 11.754 & m(4d)        & - & X \\
107  & \object{16425498-4233057} & 250.729102 & -42.551605 & -10.65 & -22.76 & 5.78 & 0.69 & 11.159 & 10.556 & 10.303 & 13.646 & m(4d)        & - & - \\
112  & \object{16442615-4218387} & 251.108983 & -42.310795 & -11.00 & -21.57 & 5.57 & 0.26 & 10.849 & 10.503 & 10.418 & 12.212 & nm           & - & - \\
116  & \object{HD 321880       } & 249.950060 & -40.311165 & -13.85 & -22.12 & 5.99 & 0.25 & 8.733 & 8.089 & 7.911 & 10.704 & nm (phot,RV) & - & - \\
119  & \object{16410838-4016078} & 250.284907 & -40.268874 & -12.03 & -20.87 & 5.53 & 0.27 & 9.924 & 9.431 & 9.248 & 11.547 & m(4d)        & - & X \\
122  & \object{16383616-3937478} & 249.650689 & -39.629952 & -12.38 & -20.76 & 5.55 & 0.25 & 9.932 & 9.331 & 9.198 & 11.467 & m(4d)        & - & - \\
123  & \object{16380672-3935178} & 249.528017 & -39.588301 & -13.72 & -21.53 & 5.86 & 0.29 & 10.152 & 9.619 & 9.407 & 11.933 & m(4d)        & - & - \\
124  & \object{16375854-3933221} & 249.493986 & -39.556155 & -14.82 & -21.77 & 6.03 & 0.35 & 10.769 & 10.069 & 9.902 & 12.929 & m(4d)        & - & - \\
125  & \object{TYC 7858-310-1  } & 249.571281 & -39.470607 & -12.16 & -21.26 & 5.62 & 0.21 & 9.125 & 8.790 & 8.711 & 10.105 & m(5d) w/RV   & - & - \\
126  & \object{HD 321933       } & 250.560323 & -39.863465 & -13.41 & -21.38 & 5.79 & 0.24 & 9.577 & 9.105 & 8.974 & 10.809 & m(5d) w/RV   & - & X \\
127  & \object{TYC 7871-176-1  } & 250.776195 & -39.812497 & -13.02 & -21.30 & 5.72 & 0.25 & 9.880 & 9.355 & 9.254 & 11.351 & m(4d)        & - & X \\
134  & \object{16532843-4057025} & 253.368481 & -40.950717 & -9.81 & -22.20 & 5.56 & 0.25 & 9.796 & 9.597 & 9.520 & 10.730 & nm           & - & - \\
137  & \object{16325443-3843525} & 248.226801 & -38.731281 & -11.21 & -21.98 & 5.65 & 0.69 & 10.679 & 10.240 & 10.073 & 12.490 & nm           & - & - \\
138  & \object{16363904-3902277} & 249.162655 & -39.041054 & -13.57 & -20.87 & 5.71 & 0.27 & 9.680 & 9.116 & 8.987 & 11.582 & m(4d)        & - & - \\
139  & \object{HD 149491       } & 249.215949 & -38.862775 & -12.61 & -21.59 & 5.74 & 0.21 & 8.946 & 8.761 & 8.634 & 9.827 & m(5d) w/RV   & H & - \\
140  & \object{HD 321858       } & 249.813745 & -39.385783 & -12.30 & -21.90 & 5.77 & 0.25 & 9.333 & 8.852 & 8.668 & 10.805 & m(4d)        & - & - \\
141  & \object{[SZB2012] 86    } & 249.947147 & -39.344561 & -11.46 & -22.04 & 5.70 & 0.81 & 10.490 & 9.753 & 9.604 & 12.754 & m(4d)        & - & X \\
142  & \object{TYC 7867-1825-1 } & 250.205341 & -39.174494 & -9.27 & -23.79 & 5.80 & 0.27 & 10.313 & 9.942 & 9.803 & 11.780 & nm           & - & - \\
143  & \object{16403359-3907217} & 250.140032 & -39.122720 & -13.63 & -20.97 & 5.73 & 0.69 & 11.402 & 10.736 & 10.507 & 13.814 & m(4d)        & - & - \\
144  & \object{16392813-3824231} & 249.867256 & -38.406428 & -11.83 & -21.33 & 5.59 & 0.26 & 9.824 & 9.352 & 9.220 & 11.233 & m(4d)        & - & X \\
148  & \object{16485929-3935423} & 252.247050 & -39.595075 & -9.65 & -22.39 & 5.57 & 0.29 & 10.092 & 9.509 & 9.311 & 11.827 & m(4d)        & - & X \\
149  & \object{16472452-3918481} & 251.852162 & -39.313364 & -10.33 & -21.67 & 5.50 & 0.26 & 9.919 & 9.439 & 9.281 & 11.375 & m(4d)        & - & X \\
151  & \object{16382553-3813591} & 249.606386 & -38.233091 & -12.90 & -22.16 & 5.88 & 0.27 & 10.219 & 9.656 & 9.461 & 12.256 & m(4d)        & - & - \\
152  & \object{HD 150641       } & 250.985442 & -38.333792 & -13.10 & -21.76 & 5.82 & 0.08 & 7.804 & 7.826 & 7.768 & 7.898 & m(4d)        & H & - \\
154  & \object{16413760-3753488} & 250.406681 & -37.896899 & -13.62 & -21.25 & 5.77 & 0.26 & 10.100 & 9.520 & 9.380 & 11.648 & m(4d)        & - & X \\
158  & \object{TYC 7868-100-1  } & 252.507688 & -37.839328 & -10.43 & -23.17 & 5.80 & 0.25 & 10.097 & 9.777 & 9.681 & 11.179 & nm           & - & - \\
159  & \object{16522966-3928195} & 253.123626 & -39.472107 & -12.02 & -20.84 & 5.51 & 0.72 & 11.069 & 10.460 & 10.323 & 13.098 & m(4d)        & - & - \\
161  & \object{TYC 7868-489-1  } & 253.812634 & -38.099953 & -11.95 & -21.80 & 5.69 & 0.24 & 9.136 & 8.522 & 8.411 & 10.906 & m(4d)        & - & X \\
164  & \object{HD 152001       } & 253.103891 & -37.607453 & -11.90 & -22.40 & 5.80 & 0.08 & 8.395 & 8.250 & 8.238 & 8.926 & m(4d)        & H & n \\
165  & \object{HD 150439       } & 250.685374 & -38.955712 & -15.06 & -21.99 & 6.09 & 0.14 & 8.036 & 7.829 & 7.745 & 8.891 & m(5d) w/RV   & H &   \\

\hline\end{tabular}
\end{sidewaystable*}

\begin{table*}[h!]
\centering
\caption{FEROS observation log. \BG{For each spectrum, we list the modified barycentric Julian Day, the integration time, the SNR at 515\,nm, the radial velocity, the bisector value if it could be measured, and the width of the CCF peak.}}
\begin{tabular}{|r|S[table-format=-5.5]|c|S[table-format=-3]|r@{$\pm$}l|r@{$\pm$}l|S[table-format=-3.2]|}
\hline
 nr  & \multicolumn{1}{c|}{MBJD} & exp.time & \multicolumn{1}{c|}{SNR} & \multicolumn{2}{c|}{RV} & \multicolumn{2}{c|}{Bisector} & \multicolumn{1}{c|}{CCFwidth} \\
  &  & (s) & \multicolumn{1}{c|}{(at 515\,nm)} & \multicolumn{2}{c|}{(km/s)} & \multicolumn{2}{c|}{} & \multicolumn{1}{c|}{($\AA$)} \\
\hline
  6  & 57952.08029 &  300 &  71 &   +3.61 &  0.21 &   -1.755 & 0.011 & 29.5 \\
 10  & 57952.13738 &  300 & 130 &   +2.40 &  0.10 &      \multicolumn{2}{c|}{---} & 23.4 \\
 13  & 57952.17475 &  300 & 127 &   +1.53 &  0.07 &      \multicolumn{2}{c|}{---} & 19.1 \\
 14  & 57952.18116 &  300 &  89 &   +1.79 &  0.02 &    0.052 & 0.010 &  8.1 \\
 15  & 57952.19783 &  300 & 108 &   +1.95 &  0.10 &      \multicolumn{2}{c|}{---} & 20.6 \\
 16  & 57952.20420 &  300 & 138 &   +3.64 &  0.31 &    1.542 & 0.008 & 28.8 \\
 17  & 57952.21987 &  300 &  39 &   +2.41 &  0.04 &      \multicolumn{2}{c|}{---}& 11.2 \\
 17  & 57952.22880 &  600 &  50 &   +2.42 &  0.03 &   -0.032 & 0.015 & 11.2 \\
 20  & 57953.11131 &  300 & 159 &   +2.49 &  0.10 &      \multicolumn{2}{c|}{---} & 24.8 \\
 21  & 57953.13064 &  300 &  79 &   +2.97 &  0.06 &      \multicolumn{2}{c|}{---} & 14.8 \\
 25  & 57953.17448 &  600 &  91 &   +1.03 &  0.02 &   -0.115 & 0.010 & 10.6 \\
 25  & 57953.16868 &  300 &  62 &   +1.11 &  0.03 &   -0.113 & 0.013 & 10.7 \\
 29  & 57953.21403 &  300 & 165 &   +2.07 &  0.36 &   -0.118 & 0.007 & 29.1 \\
 34  & 57954.14386 &  300 &  82 &  -79.66 &  0.04 &    0.047 & 0.011 &  6.1 \\
 34  & 57954.13572 &  300 &  79 &  -79.19 &  0.04 &    0.158 & 0.011 &  6.1 \\
 34  & 57954.14792 &  300 &  79 &  -79.79 &  0.04 &    0.142 & 0.011 &  6.1 \\
 34  & 57954.13980 &  300 &  82 &  -79.35 &  0.04 &    0.106 & 0.011 &  6.1 \\
 35  & 57954.25678 &  300 &  73 &   +7.02 &  6.91 &      \multicolumn{2}{c|}{---} & 46.6 \\
 35  & 57954.26084 &  300 &  74 &   -0.77 &  6.78 &    0.122 & 0.011 & 46.9 \\
 35  & 57954.25272 &  300 &  96 &   +7.60 &  5.56 &    0.352 & 0.010 & 47.5 \\
 55  & 57954.04746 &  300 & 152 &  -33.30 &  0.02 &    0.939 & 0.008 & 10.4 \\
 55  & 57954.04341 &  300 & 149 &  -33.23 &  0.02 &      \multicolumn{2}{c|}{---} & 10.5 \\
 55  & 57954.03935 &  300 & 143 &  -33.24 &  0.02 &      \multicolumn{2}{c|}{---} & 10.5 \\
 56  & 57954.09113 &  300 & 102 &   +1.74 &  0.05 &      \multicolumn{2}{c|}{---} & 14.3 \\
 56  & 57954.09521 &  300 & 105 &   +1.83 &  0.05 &      \multicolumn{2}{c|}{---} & 14.3 \\
 56  & 57954.09925 &  300 &  97 &   +1.73 &  0.05 &   -0.202 & 0.010 & 14.3 \\
116  & 57954.12032 &  300 &  52 &  -66.09 &  0.01 &    0.003 & 0.014 &  4.5 \\
116  & 57954.11217 &  300 &  53 &  -65.97 &  0.01 &   -0.148 & 0.014 &  4.6 \\
116  & 57954.11625 &  300 &  53 &  -66.06 &  0.01 &    0.054 & 0.014 &  4.5 \\
116  & 57954.10811 &  300 &  53 &  -66.08 &  0.01 &    0.093 & 0.014 &  4.5 \\
120  & 57954.22716 &  300 & 171 &  -46.60 &  0.01 &    0.055 & 0.007 &  5.2 \\
120  & 57954.22292 &  300 & 176 &  -46.60 &  0.01 &    0.042 & 0.007 &  5.2 \\
125  & 57954.08564 &  300 & 102 &   -3.18 &  0.01 &    0.025 & 0.009 &  4.4 \\
125  & 57954.08159 &  300 & 102 &   -3.15 &  0.01 &    0.070 & 0.009 &  4.4 \\
125  & 57954.07754 &  300 & 102 &   -3.14 &  0.01 &    0.077 & 0.009 &  4.4 \\
126  & 57954.18753 &  300 &  32 &   +6.43 &  0.02 &    0.102 & 0.020 &  5.9 \\
126  & 57954.17533 &  300 &  64 &   +6.21 &  0.01 &    0.135 & 0.012 &  6.0 \\
126  & 57954.17940 &  300 &  62 &   +6.29 &  0.01 &    0.129 & 0.013 &  6.1 \\
126  & 57954.18347 &  300 &  56 &   +6.22 &  0.01 &    0.069 & 0.013 &  6.1 \\
139  & 57954.06148 &  300 & 124 &   +5.13 &  0.06 &    0.126 & 0.009 & 16.6 \\
139  & 57954.05741 &  300 & 127 &   +5.10 &  0.06 &      \multicolumn{2}{c|}{---} & 16.5 \\
139  & 57954.05336 &  300 & 127 &   +5.06 &  0.06 &      \multicolumn{2}{c|}{---} & 16.5 \\
165  & 57954.20870 &  300 &  89 &   +1.62 &  0.11 &      \multicolumn{2}{c|}{---} & 20.7 \\
165  & 57954.21700 &  300 & 118 &   +1.46 &  0.08 &      \multicolumn{2}{c|}{---} & 20.8 \\
165  & 57954.21293 &  300 &  97 &   +1.51 &  0.10 &      \multicolumn{2}{c|}{---} & 20.7 \\
165  & 57954.20464 &  300 &  79 &   +1.14 &  0.11 &      \multicolumn{2}{c|}{---} & 20.6 \\

\hline\end{tabular}
\label{FEROSlog}
\end{table*}
\begin{table*}[h!]
\centering
\caption{\label{FEROSmean}Astrophysical parameters of member stars in the V1062 Sco moving group, \BG{followed by 4 rejected candidates. The measurements are weighted averages when multiple spectra were observed. In this case the uncertainties are the error on the mean. RV from the CERES pipeline, stellar parameters from ZASPE.}}
\begin{tabular}{|r|r@{$\pm$}l|c|r|r|r|} %
\hline
 nr & \multicolumn{2}{c|}{RV} & $T_{\rm eff}$ & $\log g$ & $v\sin i$ & [Fe/H] \\
  & \multicolumn{2}{c|}{(km/s)} & (K) &  & (km/s) &  \\
\hline
  6  &   +3.61 &  0.21 &  6026  & 3.8  &  51.8  & +0.22  \\
 10  &   +2.40 &  0.10 &  6295  & 3.8  &  38.8  & -0.02  \\
 13  &   +1.53 &  0.07 &  6406  & 3.9  &  31.5  & +0.08  \\
 14  &   +1.79 &  0.02 &  5772  & 4.1  &  12.3  & +0.04  \\
 15  &   +1.95 &  0.10 &  6387  & 3.8  &  33.8  & +0.04  \\
 16  &   +3.64 &  0.31 &  6702  & 3.8  &  66.8  & +0.10  \\
 17  &   +2.41 &  0.02 &  5032  & 4.0  &  17.0  & +0.16  \\
 20  &   +2.49 &  0.10 &  5882  & 2.5  &  37.1  & -0.52  \\
 21  &   +2.97 &  0.06 &   ---  & ---  &   ---  &   ---  \\
 25  &   +1.06 &  0.02 &  5332  & 4.2  &  16.7  & +0.05  \\
 29  &   +2.07 &  0.36 &  6197  & 4.0  &  79.2  & -0.04  \\
 56  &   +1.77 &  0.03 &  6144  & 4.0  &  23.5  & +0.04  \\
125  &   -3.16 &  0.01 &  6254  & 4.3  &   4.9  & +0.03  \\
126  &   +6.26 &  0.01 &  5336  & 4.2  &   8.1  & +0.03  \\
139  &   +5.10 &  0.03 &  6383  & 3.9  &  27.5  & +0.01  \\
165  &   +1.45 &  0.05 &  6204  & 3.9  &  33.2  & +0.08  \\
 35  &     \multicolumn{2}{c|}{---} &  6020  & 3.5  & 119.9  & +0.15  \\
\hline
 34  &  -79.50 &  0.02 &  6790  & 3.6  &   8.8  & -0.52  \\
 55  &  -33.26 &  0.01 &  6508  & 4.3  &  17.0  & +0.17  \\
116  &  -66.05 &  0.01 &  5022  & 2.7  &   5.0  & -0.37  \\
120  &  -46.60 &  0.00 &   ---  & ---  &   ---  &   ---  \\
 
\hline\end{tabular}
\end{table*}

\end{appendix}
\end{document}